\documentclass[utf8]{frontiersFPHY} 

\newcommand{\R}[1]{\textcolor{red}{#1}}

\newcommand{\be}{\begin{equation}}
\newcommand{\ee}{\end{equation}}
\newcommand{\bea}{\begin{eqnarray}}
\newcommand{\eea}{\end{eqnarray}}
\def \blank{\mbox{}}

\def\x{\mbox{{\bf x}}}
\def\q{\mbox{{\bf q}}}

\def\R{\mbox{{\bf R}}}

\def\X{\mbox{{\bf X}}}

\def\Y{\mbox{$\bf{Y}$}}
\def\y{\mbox{$\bf{y}$}}

\def\f{\mbox{$\bf{f}$}}
\def\F{\mbox{$\bf{F}$}}

\usepackage{graphicx}
\usepackage{xcolor}

\usepackage{url,hyperref,lineno,microtype,subcaption}
\usepackage[onehalfspacing]{setspace}

\def\keyFont{\fontsize{8}{11}\helveticabold }
\def\firstAuthorLast{Miller {et~al.}} 
\def\Authors{Anna Miller\,$^{1,*}$, Dawei Li\,$^{1}$ Jason Platt\,$^{1}$ Arij Daou\,$^{4}$ Daniel Margoliash\,$^{2}$ and Henry D. I. Abarbanel\,$^{1,3}$}
\def\Address{$^{1}$Department of Physics, University of California San Diego, La Jolla , CA, USA \\
$^{2}$ Department of Anatomy and Organismal Biology, University of Chicago, Chicago , IL, USA \\
$^{3}$ Marine Physical Laboratory (Scripps Institution of Oceanography) and Department of Physics, University of California, La Jolla, CA, USA  \\
$^{4}$ Department of Biomedical Engineering, American University of Beirut, Beirut, Lebanon }
\def\corrAuthor{Anna Miller}

\def\corrEmail{a8miller@ucsd.edu}

\begin{document}
\onecolumn
\firstpage{1}

\title[]{Statistical Data Assimilation: Formulation and Examples from Neurobiology}

\author[\firstAuthorLast]{\Authors}
\address{}
\correspondance{}

\extraAuth{}

\maketitle

\begin{abstract}
\section{}
For the Research Topic {\em Data Assimilation and Control: Theory and Applications in Life Sciences} we first review the formulation of statistical data assimilation (SDA) and discuss algorithms for exploring variational approximations to the conditional expected values of biophysical aspects of functional neural circuits. Then we report on the application of SDA to (1) the exploration of properties of individual neurons in the HVC nucleus of the avian song system, and (2) characterizing individual neurons formulated as very large scale integration (VLSI) analog circuits with a goal of building functional, biophysically realistic, VLSI representations of functional nervous systems. Networks of neurons pose a substantially greater challenge, and we comment on formulating experiments to probe the properties, especially the functional connectivity, in song command circuits within HVC.

\tiny
 \keyFont{ \section{Keywords:} Data Assimilation, Neuronal Dynamics, HVC, Ion Channel Properties, Variational Annealing, Neuromorphic, VLSI}
 
\end{abstract}

\section{Introduction}

A broad class of `inverse' problems presents itself in many scientific and engineering inquiries. The overall question addressed by these is how to transfer information from laboratory and field observations to candidate models of the processes underlying those observations. 

The existence of large, information rich, well curated data sets from increasingly sophisticated observations on complicated nonlinear systems has set new challenges to the information transfer task. Assisting with this challenge are new substantial computational capabilities. 

Together they have provided an arena in which principled formulation of this information transfer along with algorithms to effect the transfer have come to play an essential role. This paper reports on some efforts to meet this class of challenge within neuroscience. Many of the ideas are applicable much more broadly than our focus, and we hope the reader will find this helpful in their own inquiries.

In this special issue entitled {\em Data Assimilation and Control: Theory and Applications in Life Sciences}, of the journal {\em Frontiers in Applied Mathematics and Statistics--Dynamical Systems}, we participate in the broader quantitative setting for this information transfer. The procedures are called ``data assimilation'' following its use in the effort to develop realistic numerical weather prediction models~\citep*{lor96,even} over many decades. We prefer the term `statistical data assimilation' (SDA) to emphasize that key ingredients in the procedures involved in the transfer rest on noisy data and on recognizing errors in the models to which information in the noisy data is to be transferred.

This article begins with a formulation of SDA with some additional clarity beyond the discussion in~\citep{abar13}. We also discuss some algorithms helpful for implementing the information transfer, testing model compatibility with the available data, and quantitatively identifying how much information in the data can be represented in the model selected by the SDA user. Using SDA will also remind us that data assimilation efforts are well cast as problems in statistical physics~\citep{tong2011}.

After the discussion of SDA, we turn to some working ideas on how to perform the high dimensional integrals involved in SDA. In particular we address the `standard model' of SDA where data is contaminated by Gaussian noise and model errors are represented by Gaussian noise, though the integrals to be performed are, of course, not Gaussian. The topics include the approximation of Laplace~\cite{Laplace1774,Laplace1986} and Monte Carlo methods.

With these tools in hand, we turn to neurobiological questions that arise in the analysis of individual neurons and, in planning, for network components of the avian song production pathway. These questions are nicely formulated in the general framework, and we dwell on specifics of SDA in a realistic biological context. The penultimate topic we address is the use of SDA to calibrate VLSI analog chips designed and built as components of a developing instantiation of the full songbird song command network, called HVC. Lastly, we discuss the potential utlization of SDA for exploring biological networks.

At the outset of this article we may expect that our readers from Physics and Applied Mathematics along with our readers from Neurobiology may find the conjunction of the two ``strange bedfellows'' to be incongruous. For the opportunity to illuminate the natural melding of the facets of both kinds of questions, we thank the editors of this special issue.

\section{Materials and Methods}
\subsection{General Overview of Data Assimilation}

We will provide a structure within which we will frame our discussion of transfer of information from data to a model of the underlying processes producing the data.

We start with a observation window in time $[t_0,t_F]$ within which we make a set of measurements at times $ t =   \{\tau_1, \tau_2, ...,\tau_k, ..., \tau_F\};\;t_0 \le \tau_k \le t_F$. At each of these measurement times, we observe $L$ quantities $\y(\tau_k) = \{y_1(\tau_k), y_2(\tau_k), ..., y_L(\tau_k)\}$. The number $L$ of observations at each measurement time $\tau_k$ is typically less, often much less, than the number of degrees of freedom $D$ in the observed system; $D \gg L$.

These are views into the dynamical processes of a system we wish to characterize. The quantitative characterization is through a model we choose. It describes the interactions among the states of the observed system. If we are observing the time course of a neuron, for example, we might measure the membrane voltage $y_1(\tau_k) = V_m(\tau_k)$ and the intracellular Ca$^{2+}$ concentration $y_2(\tau_k) = [Ca^{2+}](\tau_k)$. From these data we want to estimate the unmeasured states of the model as a function of time as well as estimate biophysical parameters in the model.

The processes characterizing the state of the system (neuron) we call $x_a(t);\;a=1,2,...,D \ge L$, and they are selected by the user to describe the dynamical behavior of the observations through a set of equations in continuous time
\be 
\frac{dx_a(t)}{dt} = F_a(\x(t),\q),
\label{dynamics}
\ee
or in discrete time $t_n = t_0 + n\Delta t;\;n=0,1,...,N;\;t_N = t_F$ via
\be
x_a(t_{n+1}) = x_a(n+1) = f_a(\x(t_n),\q) = f_a(\x(n),\q),
\label{discrete}
\ee 
where $\q$ is a set of fixed parameters associated with the model. $\f(\x(n),\q)$ is related to $\F(\x(t),\q)$ through the choice the user makes for solving the continuous time flow for $\x(t)$ through a numerical solution method of choice~\cite{press}.

Considering neuronal activity, Eq. (\ref{dynamics}) could be coupled Hodgkin-Huxley (HH) equations~\cite{jwu,willshaw} for voltage, ion channel gating variables, constituent concentrations, and other ingredients. If the neuron is isolated {\em in vitro}, such as by using drugs to block synaptic transmission, then there would be no synaptic input to the cell to describe. While if it is coupled to a network of neurons, their functional connectivity would be described in $\F(\x(t),\q)$ or $\f(\x(n),\q)$. Typical parameters might be maximal conductances of the ion channels, reversal potentials, and other time-independent numbers describing the kinetics of the gating variables. In many experiments $L$ is only 1, namely, the voltage across the cell membrane, while $D$ may be on the order of 100; Hence $D \gg L$.

As we proceed from the initiation of the observation window at $t_0$ we must move our model equation variables $\x(0)$, Eq. (\ref{discrete}), from $t_0$ to $\tau_1$ where a measurement is made. Then using the model dynamics we move along to $\tau_2$ and so forth until we reach the last measurement time $\tau_F$ and finally move the model from $\x(\tau_F)$ to $\x(t_F)$. In each stepping of the model equations Eq. (\ref{discrete}) we may make many steps of $\Delta t$ in time to achieve accuracy in the representation of the model dynamics. The full set of times $t_n$ at which we evaluate the model $\x(t_n)$ we collect into the path of the state of the model through $D$-dimensional space: $\X = \{\x(0),\x(1),...,\x(n),...,\x(N) = \x(F)\}$. The dimension of the path is $(N+1)D + N_q$, where $N_q$ is the number of parameters $\q$ in our model.

\begin{figure}[tbph] 
  \centering
   \includegraphics[width=180mm,keepaspectratio]{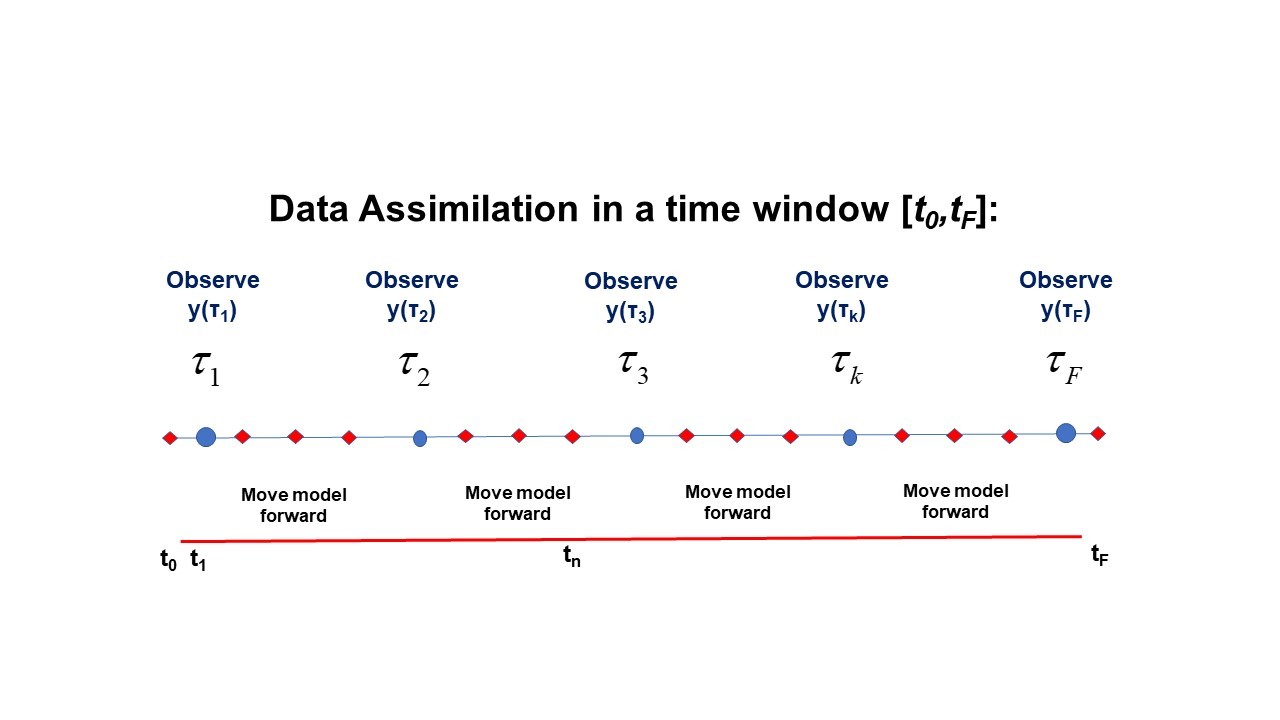}
  \caption{A visual representation of the window $t_0 \le t \le t_F$ in time during which $L$-dimensional observations $\y(\tau_k)$ are performed at observation times $t = \tau_k;\;k=1,...,F$. This also shows times at which the $D$-dimensional model developed by the user $\x(n+1) = \f(\x(n),\q)$ is used to move forward from time $n$ to time $n+1$: $t_n = t_0 + n\Delta t;\;n=0,1,...,N$. $D \ge L$. The path of the model $\X = \{\x(0),\x(1),...,\x(n),...,\x(N) = \x(F)\}$ and the collection $\Y$ of $L$-dimensional observations at each observation time $\tau_k$, $\Y = \{\y(\tau_1),\y(\tau_2),...,\y(\tau_k),...,\y(\tau_F\}$ ($\y = \{y_1, y_2, ..., y_L\}$) is also indicated. }
  \label{dataassim1}
\end{figure}

It is worth a pause here to note that we have now collected two of the needed three ingredients to effect our transfer of the information in the collection of all measurements $\Y = \{\y(\tau_1),\y(\tau_2),...,\y(\tau_F)\}$ to the model $\f(\x(n),\q)$ along the path $\X$ through the observation window $[t_0,t_F]$: (1) data $\Y$ and (2) a model of the processes in $\Y$, devised by our experience and knowledge of those processes. The notation and a visual presentation of this is found in Fig. (\ref{dataassim1}).

The {\bf third} ingredient, comprised of methods to generate the transfer from $\Y$ to properties of the model, will command our attention throughout most of this paper. If the transfer methods are successful and, according to some metric of success, we arrange matters so that at the measurement times $\tau_k$, the $L$ model variables $\x(t)$ associated with $\y(\tau_k)$ are such that $x_l(\tau_k) \approx y_l(\tau_k)$, we are {\em not} finished. We have then only demonstrated that the model is consistent with the known data $\Y$. We must use the model, completed by the estimates of $\q$ and the state of the model at $t_F, \;\x(t_F)$, to predict forward for $t > t_F$, and we should succeed in comparison with measurements for $\y(\tau_r)$ for $\tau_r > t_F$. As the measure of success of predictions, we may use the same metric as utilized in the observation window.

As a small aside, the same overall setup applies to supervised machine learning networks~\cite{abar18} where the observation window is called the training set; the prediction window is called the test set, and prediction is called generalization.

\subsubsection{The Data are Noisy; the Model has Errors}

Inevitably, the data we collect is noisy, and equally the model we select to describe the production of those data has errors. This means we must, at the outset, address a conditional probability distribution $P(\X|\Y)$ as our goal in the data assimilation transfer from $\Y$ to the model. In~\cite{abar13} we describe how to use the Markov nature of the model $\x(n) \to \x(n+1) = \f(\x(n),\q)$ and the definition of conditional probabilities to derive the recursion relation:
\bea
{\textcolor {blue}{P(\X(n+1)|\Y(n+1))}} &=& \frac{P(\y(n+1),\x(n+1),\X(n)|\Y(n))}{P(\y(n+1)|\Y(n))\,P(\x(n+1),\X(n+1)|\Y(n))} \bullet \nonumber \\
&&P(\x(n+1)|\x(n)) {\textcolor {blue} {P(\X(n)|\Y(n))}} \nonumber \\
&=& \exp[CMI(\y(n+1),\x(n+1),\X(n)|\Y(n))] \bullet \nonumber \\
&&P(\x(n+1)|\x(n)) {\textcolor {blue} {P(\X(n)|\Y(n))}},
\eea
where we have identified $CMI(a,b|c) = \log\left[\frac{(P(a,b|c)}{P(a|c)\,P(a|c)}\right]$. This is Shannon's conditional mutual information~\cite{fano} telling us how many bits (for $\log_2$) we know about $a$ when observing $b$ conditioned on $c$. For us $a = \{\y(n+1)\}, b = \{\x(n+1),\X(n+1)\}, c = \{\Y(n)\}$.

Using this recursion relation to move backwards through the observation window from $t_F = t_0 + N\Delta t$ through the measurements at times $\tau_k$ to the start of the window at $t_0$, we may write, up to factors independent of $\X$
\be 
P(\X|\Y) = \biggl\{\prod_{k=1}^F P(\y(\tau_k)|\X(\tau_k))\,\prod_{n=0}^{N-1}P(\x(n+1)|\x(n))\biggr\}P(\x(0)).
\ee
If we now write $P(\X|\Y) \propto \exp[-A(\X)]$ where $A(\X)$, the negative of the log likelihood, we call the action, then conditional expected values for functions along the path $\X$ are defined by
\be 
E[G(\X)|\Y] = \langle G(\X) \rangle = \frac{\int d\X\, G(\X) e^{-A(\X)}}{ \int d\X\, e^{-A(\X)}},
\label{expect}
\ee
$d\X = \prod_{n=0}^N\,d^D\x(n)$, and all factors in the action independent of $\X$ cancel out here.
The action takes the convenient expression
\bea
A(\X) = -\biggl \{\sum_{k=1}^F\log[P(\y(\tau_k)|\X(\tau_k)] + \sum_{n=0}^{N-1} \log[P(\x(n+1)|\x(n))] \biggr\} - \log[P(\x(0))],
\eea
which is the sum of the terms which modify the conditional probability distribution when an observation is made at $t = \tau_k$ and the sum of the stochastic version of $\x(n) \to  \x(n+1) - \f(\x(n),\q)$ and finally the distribution when the observation window opens at $t_0$.

What quantities $G(\X)$ are of interest? One natural one is the path $G(\X) = \X_{\mu}; \mu = \{a,n\}$ itself; another is the covariance around that mean $\langle \X_{\mu}\rangle = \bar{\X}_{\mu} = \langle \X_{\mu} \rangle: \langle (\X_{\mu} - \bar{\X}_{\mu})(\X_{\nu} - \bar{\X}_{\nu}) \rangle$. Other moments are of interest, of course. If one has an anticipated form for the distribution at large $\X$, then $G(\X)$ may be chosen as a parametrized version of that form and those parameters determined near the maximum of $P(\X|\Y)$.

The action simplifies to what we call the `standard model' of data assimilation when (1) observations $\y$ are related to their model counterparts via Gaussian noise with zero mean and diagonal precision matrix $\R_m$, and (2) model errors are associated with Gaussian errors of mean zero and diagonal precision matrix $\R_f$:
\be
A(\X) = \sum_{k=1}^F \sum_{l=1}^L \frac{R_m(k)}{2} (x_l(\tau_k) - y_l(\tau_k))^2  + \sum_{n=0}^{N-1} \sum_{a=1}^D \frac{R_f(a)}{2} (x_a(n+1) - f_a(\x(n),\q))^2.
\label{standard}
\ee
If we have knowledge of the distribution $P(\x(0))$ at $t_0$ we may add it to this action. If we have no knowledge of $P(\x(0))$, we may take its distribution to be uniform over the dynamic range of the model variables, then it, as here, is absent, canceling numerator and denominator in Eq. (\ref{expect}). 

Our challenge is to perform integrals such as Eq. (\ref{expect}). One should anticipate that the dominant contribution to the expected value comes from the maxima of $P(\X|\Y)$ or, equivalently the minima of $A(\X)$. At such minima, the two contributions to the action, the measurement error and the model error, balance each other to accomplish the explicit transfer of information from the former to the latter.

We note, as before, that when $\f(\x(n),\q)$ is nonlinear in $\X$, as it always is in interesting examples, the expected value integral Eq. (\ref{expect}) is not Gaussian. So, some thinking is in order to approximate this high dimensional integral. We turn to that now. After consideration of methods to do the integral, we will return to a variety of examples taken from neuroscience.

The two generally useful methods available for evaluating this kind of high dimensional integral are Laplace's method~\cite{Laplace1774,Laplace1986} and Monte Carlo techniques~\cite{press,biocyb2,neal2012}. We address them in order. We also add our own new and useful versions of the methods.

\subsubsection{Laplace's Method}

To locate the minima of the action $A(\X) = - \log[P(\X|\Y)]$ we must seek paths $\X^{(j)};\;j=0,1,...$ such that $\partial A(\X)/\partial \X|_{\X^{(j)}} = 0$, and then check that the second derivative at $\X^{(j)}$, the Hessian, is a positive definite matrix in path coordinates. The vanishing of the derivative is a necessary condition. 

Laplace's method~\cite{Laplace1774} expands the action around the $\X^{(j)}$ seeking the path $\X^{(0)}$ with the smallest value of $A(\X)$. The contribution of $\X^{(0)}$ to the integral Eq. (\ref{expect}) is approximately $\exp[A(\X^{(1)}) -  A(\X^{(0)})]$ bigger than that of the path with the next smallest action. 

This sounds more or less straightforward; however, finding the global minimum of a nonlinear function such as $A(\X)$ is an NP-complete 
problem~\cite{murty87}. In a practical sense one cannot expect to succeed with such problems. However there is an attractive feature of the form of 
$A(\X)$ that permits us to accomplish more.

We now discuss two algorithmic approaches to implementing Laplace's method.

\subsubsection{Precision Annealing for Laplace's Method}

Looking at Eq. (\ref{standard}) we see that if the precision of the model is zero, $R_f = 0$, the action is quadratic in the $L$ measured variables $x_l(n)$ and independent of the remaining states. The global minimum of such an action comes with $x_l(\tau_k) = y_l(\tau_k)$ and any choice for the remaining states and parameters. Choose the path with these values of $\x(\tau_k)$ and values from a uniform distribution of the other state variables and parameters covering the expected dynamic range of those, and call it path $\X_{\mbox{init}}$. In practice, we recognize that the global minimum of $A(\X)$ is degenerate at $R_f = 0$, so we select many initial paths. We choose $N_I$ of them, and initialize whatever numerical optimization program we have selected, to run on each of them. We continue to call the collection of $N_I$ paths $\X_{\mbox{init}}$. 

\begin{itemize} 
  \item Now we increase $R_f$ from $R_f = 0$ to a small value $R_{f0}$. Use each of the $N_I$ paths in $\X_{\mbox{init}}$ as initial conditions for our numerical optimization program chosen to find the minima of $A(\X)$, and we arrive at $N_I$ paths $\X_0$. Evaluate $A(\X_0)$ on all $N_I$ paths $\X_0$.
  \item We increase $R_f = R_{f0} \to R_{f0}\alpha;\;\alpha > 1$, and now use the $N_I$ paths $\X_0$ as the initial conditions for our numerical optimization program chosen to find the minima of $A(\X)$, we arrive at $N_I$ paths $\X_1$. Evaluate $A(\X_1)$ on all $N_I$ paths $\X_1$.
  \item We increase $R_f = R_{f0} \alpha \to R_{f0}\alpha^2$. Now use the $N_I$ paths $\X_1$ as the initial conditions for our numerical optimization program chosen to find the minima of $A(\X)$, we arrive at $N_I$ paths $\X_2$. Evaluate $A(\X_2)$ on all $N_I$ paths $\X_2$.
  \item Continue in this manner increasing $R_f$ to $R_f = R_{f0}\alpha^{\beta};\;\beta = 0,1,...$, then using the selected numerical optimization program to arrive at $N_I$ paths $\X_{\beta}$.  Evaluate $A(\X_{\beta})$ on all $N_I$ paths $\X_{\beta}$.
  \item As a function of $\log_{\alpha}\left[\frac{R_f}{R_{f0}}\right]$ display all $N_I$ values of $A(\X_{\beta})$ versus $\beta$ for all $\beta = 0,1,2,...\beta_{\mbox{max}}$.
\end{itemize} 

We call this method {\em precision annealing} (PA)~\cite{Ye-et-al,Ye2015,ye,quinn}. It slowly turns up the precision of the model collecting paths at each $R_f$ that emerged from the degenerate global minimum at $R_f = 0$. In practice it is able to track $N_I$ possible minima of $A(\X)$ at each $R_f$. When not enough information is presented to the model, that is $L$ is too small, there are many local minima at all $R_f$. This is a manifestation of the NP-completeness of the minimization of $A(\X)$ problem. None of the minima may dominate the expected value integral of interest. 

As $L$ increases, and enough information is transmitted to the model, for large $R_f$ one minimum appears to stand out as the global minimum, and the paths associated with that smallest minimum yields good predictions. We note that there are always paths, not just a single path, as we have a distribution of paths, $N_I$ of which are sampled in the PA procedure, within a variation of size $1/\sqrt{R_m}$. A clear example of this is seen in~\cite{sasha} in a small, illustrative model.

\subsubsection{``Nudging'' within Laplace's Method}

In meteorology one approach to data assimilation is to add a term to the deterministic dynamics which move the state of a model towards the observations~\cite{anthes1974}
\be 
x_a(n+1) = f_a(\x(n),\q) + u(n)(y_l(n) - x_l(n))\delta_{al},
\label{nudging}
\ee
where $u(n) > 0$ and vanishes except where a measurement is available. This is referred to as `nudging'. It appears in an {\em ad hoc}, but quite useful, manner.

Within the structure we have developed, one may see that the `nudging term' arises through the balance between the measurement error term and the model error term in the action. This is easy to see when we look at the continuous time version of the data assimilation standard model
\bea 
&&A(\x(t),\dot{\x}(t)) = \int_{t_0}^{t_F} dt\, \biggl \{ \sum_{l=1}^{L}\, \frac{R_m(t,l)}{2}(x_l(t) - y_l(t))^2 \nonumber \\
&&+ \sum_{a=1}^D \frac{R_f(a)}{2} (\dot{x}_a(t) - F_a(\x(t),\q))^2 \biggr \}.
\eea

The extremum of this action is given by the Euler-Lagrange equations for the variational problem~\cite{gfomin}
\bea
&&\biggl[\delta_{ab} \frac{d \blank }{dt} + \frac{\partial F_b(\x(t)}{\partial x_a(t)}\biggr]\biggl[ \dot{x}_b(t) - F_b(\x(t))\biggr] \nonumber \\
&&= \frac{R_m(a,t)}{R_f(a)}\delta_{al}(x_l(t) - y_l(t)),
\eea
in which the right hand side is the `nudging' term appearing in a natural manner. Approximating the operator $\delta_{ab} \frac{d \blank }{dt} + \frac{\partial F_b(\x(t)}{\partial x_a(t)}$
we can rewrite this Euler-Lagrange equation in `nudging' form
\be 
\frac{dx_a(t)}{dt} = F_a(\x(t)) + u(t)\delta_{al}(x_l(t) - y_l(t)).
\ee

We will  use both the full variation of the action, in discrete time, as well as its nudging form in our examples below.

\subsubsection{Monte Carlo Methods}

Monte Carlo methods~\cite{MC,biocyb2,quinn,press} are well covered in the literature. We have not used them in the examples in this paper. However, the development of a precision annealing version of Monte Carlo techniques promises to address the difficulties with large matrices for the Jacobian and Hessians required in variational principles~\cite{wong2018}. When one comes to network problems, about which we comment later, this method may be essential.

\section{Results}
\subsection{Using SDA to Analyze the Avian Song System}

We take our examples of the use of SDA in neurobiology from experiments on the avian song system. These have been performed in the University of Chicago laboratory of Daniel Margoliash, and we do not plan to describe in any detail the experiments nor the avian song production pathways in the avian brain. We give the essentials of the experiments and direct the reader to our references to develop the full biologically oriented idea why this system is enormously interesting.

Essentially, however, the manner in which songbirds learn and produce their functional vocalization---song---is an elegant non-human example of a behavior that is cultural: the song is determined both by a genetic substrate and, interestingly, by refinement on top of that substrate by juveniles learning the song from their (male) elders. The analogs to learning speech  in humans~\cite{DoupeKuhl99} are striking.

Our avian singer is a zebra finch. They, as do most other songbirds, learn vocal expression through auditory feedback~\cite{Daou2013,natrevneuro11,songbirdlearning,Mooney2009,DoupeKuhl99}.  This makes the study of the song system a good model for learning complex behavior~\cite{Simonyan2012,songbirdlearning,Teramitsu2004}. Parts of the song system are analogous to the mammalian basal ganglia and regions of the frontal cortex~\cite{songbirdlearning,Jarvis2005,Doupe2005}.  Zebra finch in particular have the attractive property of singing only a single learned song, and with high precision, throughout their adult life.

Beyond the auditory pathways themselves, two neural pathways are principally responsible for song acquisition and production in zebra finch. The first is the Anterior Forebrain Pathway (AFP) which modulates learning.  The second is a posterior pathway responsible for directing song production: the Song Motor Pathway (SMP)~\cite{natrevneuro11,Mooney2009,Nottebohm76}. The HVC nucleus in the avian brain uniquely contributes to both of these~\cite{Mooney2009}.

\begin{figure}[bp] 
  \centering
  \includegraphics[width=5.67in,keepaspectratio]{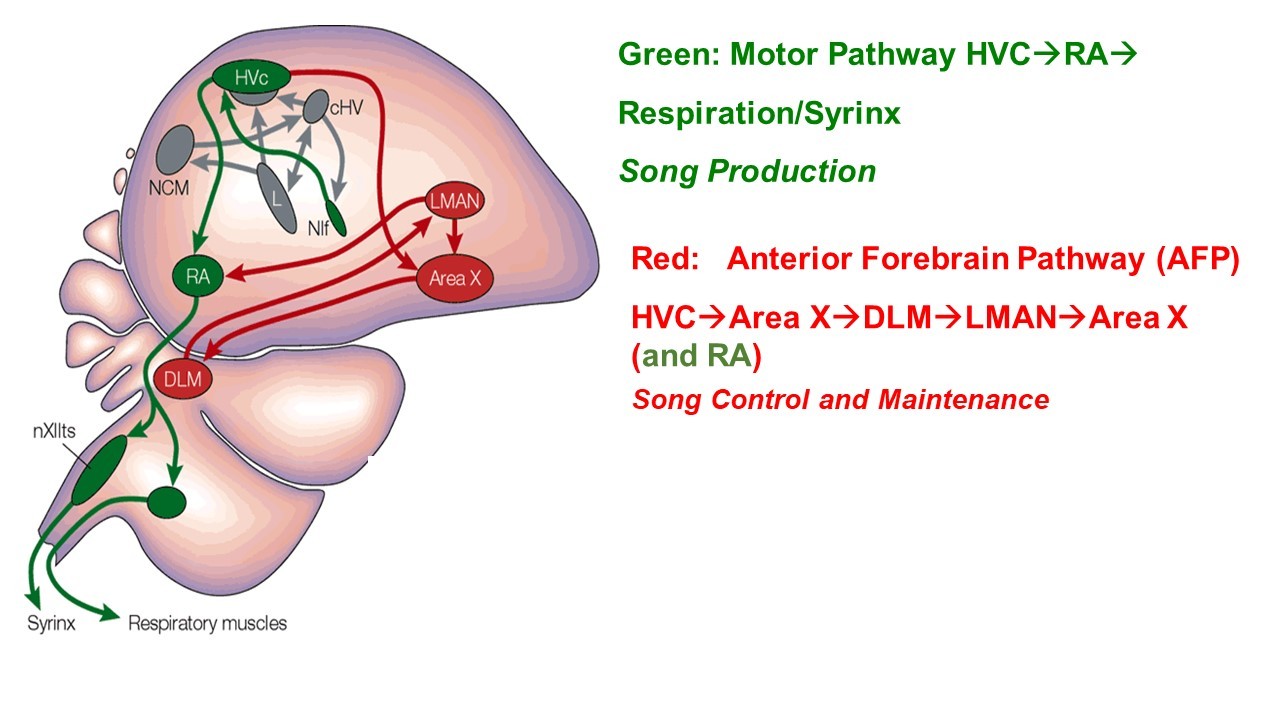}
  \caption{A drawing of the Song Production Pathway and the Anterior Forebrain Pathway of avian songbirds. Parts of the auditory pathways are shown in grey. Pathways from the brainstem that ultimately return to HVC are not shown.
\\ 
The HVC network image is taken from~\citep{doupe2000}. Reprinted by permission from Copyright Clearance Center: Springer Nature, \textit{Nature Reviews Neuroscience}, Auditory Feedback in Learning and Maintenance of Vocal Behavior, M. S. Brainard and A. J. Doupe, 2000.}
  \label{HVCbrain_doupe}
\end{figure}

There are two principal classes of projection neurons which extend from HVC: neurons which project to the robust nucleus of the arcopallium ($\text{HVC}_{RA}$), and neurons which project to  Area X ($\text{HVC}_X$). HVC$_{RA}$ neurons extend to the SMP pathway and HVC$_X$ neurons extend to the AFP~\cite{Mooney2009,Margoliash95}. These two classes of projection neurons combined with classes of HVC interneurons, make up the three broad classes of neurons within HVC. Fig. (\ref{HVCbrain_doupe})~\cite{doupe2000} displays these structures in the avian brain.

{\em In vitro} observations of each HVC cell type have been obtained through patch-clamp techniques making intracellular voltage measurements in a reduced, brain slice preparation~\cite{Daou2013}. In this configuration, the electrode can simultaneously inject current into the neuron while measuring the whole cell voltage response~\cite{Hamill1981}. From these data, one can establish the physical parameters of the system~\cite{Daou2013}. Traditionally this is done using neurochemicals to block selected ion channels and measuring the response properties of others~\cite{Hernandez2012}. Single current behavior is recorded and parameters are found through mathematical fits of the data. This procedure has its drawbacks, of course. There are various technical problems with the choice of channel blockers. Many of even the modern channel blockers are not subtype specific~\cite{Bagal2015} and may only partially block channels~\cite{frolov}. A deeper conceptual problem is that is difficult to know what channels one may have missed altogether. Perhaps there are channels which express themselves only outside the bounds of the experimental conditions.

Our solution to such problems is the utilization of statistical data assimilation (SDA). This a method developed by meteorologists and others as computational models of increasingly large dynamical systems have been desired. Data assimilation has been described in our earlier sections. 

In this paper, we focus on the song learning pathway, reporting on experiments involving the HVC$_X$ neuron. The methods are generally applicable to the other neurons in HVC, and actually, to neurons seen as dynamical systems in general.

We start with a discussion about the neuron model. First we demonstrate the utility of our precision annealing methods through the use of {\it twin experiments}. These are numerical experiments in which `data' is generated through a known model (of HVC$_X$), then analyzed via precision annealing. In a twin experiment, we know everything, so we can verify the SDA method by looking at predictions after a observation window in which the model is trained, and we may also compare the estimations of {\em unobserved state variables and parameters} to the ingredients and output of the model.

Twin experiments are meant to mirror the circumstances of the real experiment. We start by taking the model that we think describes our physical system. Initial points for the state variables and parameters are chosen, which are used along with the model to numerically integrate forward in time. This leaves us with complete information about the system. Noise is added to a subset of the state variables to emulate the data to be collected in a lab experiment. We then perform PA on these simulated data, as if they were real data. The results of these numerical experiments can be used to inform laboratory experiments, and indeed help design them, by identifying the necessary measurements and stimulus needed to accurately electrophysiologically characterize a neuron.

The second set of SDA analyses we report on using `nudging', as described above, to estimate some key biophysical properties of HVC$_X$ neurons from laboratory data. This SDA procedure is applied to HVC$_X$ neurons in two different birds. The results show that though each bird is capable of normal vocalization, their complement of ion channel strengths is apparently different. We report on a suggestive example of this, leaving a full discussion to~\cite{margdaou}.

In order to obtain good estimation results, we must choose a forcing or stimulus with the model in mind: the dynamical range of the neuron must be thoroughly explored. This suggests a few key properties of the stimulus:

\begin{itemize}
\item The current waveform of $I_{injected}(t)$ must have sufficient amplitudes ($\pm$) and must be applied sufficiently long in time that it explores the full range of the neuron variation.
\item The frequency content of the stimulus current must be a low enough that it does not exceed the low-pass cutoff frequency associated with the RC time constant of the neuron. This cutoff is typically in the neighborhood of 50-100Hz.
\item The current must explore all time scales expressed in the neuron's behavior. 
\end{itemize}

\subsection{Analysis of HVC$_X$ Data}

The model for an HVC$_X$ neuron is substantially taken from~\cite{Daou2013} and described in our Appendix. We now use this model in a `twin experiment' in which PA is utilized, and then using `nudging' we present the analysis of experimental data on two Zebra Finch.

\subsubsection{Twin Experiment on HVC$_X$ Neuron Model}

A twin experiment is a synthetic numerical experiment meant to mirror the conditions of a laboratory experiment. We use our mathematical model with some informed parameter choices in order to generate numerical data. Noise is added to observable variables in the model, here $V(t)$. These data are then put through our SDA procedure to estimate parameters and unobserved states of the model. The neuron model is now calibrated or completed.

Using the parameters and the full state of the model at the end $t_F$ of an observation window $[t_0,t_F]$, we take a current waveform $I_{injected}(t \ge t_F)$ to drive the model neuron and predict the time course of all dynamical variables in the prediction window $[t_F,...]$. This validation of the model is the critical test of our SDA procedure, here PA. In a laboratory experiment we have no specific knowledge of the parameters in the model and, by definition, cannot observe the unobserved state variables; here we can do that. So, `fitting' the observed data within the observation window $[t_0,t_F]$ is not enough, we must reproduce all states for $t \ge t_F$ to test our SDA methods.

We use the model laid out in the Appendix. We assume that the neuron has a resting potential of $-70$ mV and set the initial values for the voltage and each gating variable accordingly. We assume that the internal calcium concentration of the cell is $C_{in} = 0.1\ \mu M$. We use an integration time step of $0.02$ ms and integrate forward in time using an adaptive Runge-Kutta method~\cite{press}. Noise is added to the voltage time course by sampling from a Gaussian distribution $\mathcal{N}(0,2)$ in units of mV. 

\begin{figure}[htbp] 
  \centering
  \includegraphics[width=5.67in,height=4.34in,keepaspectratio]{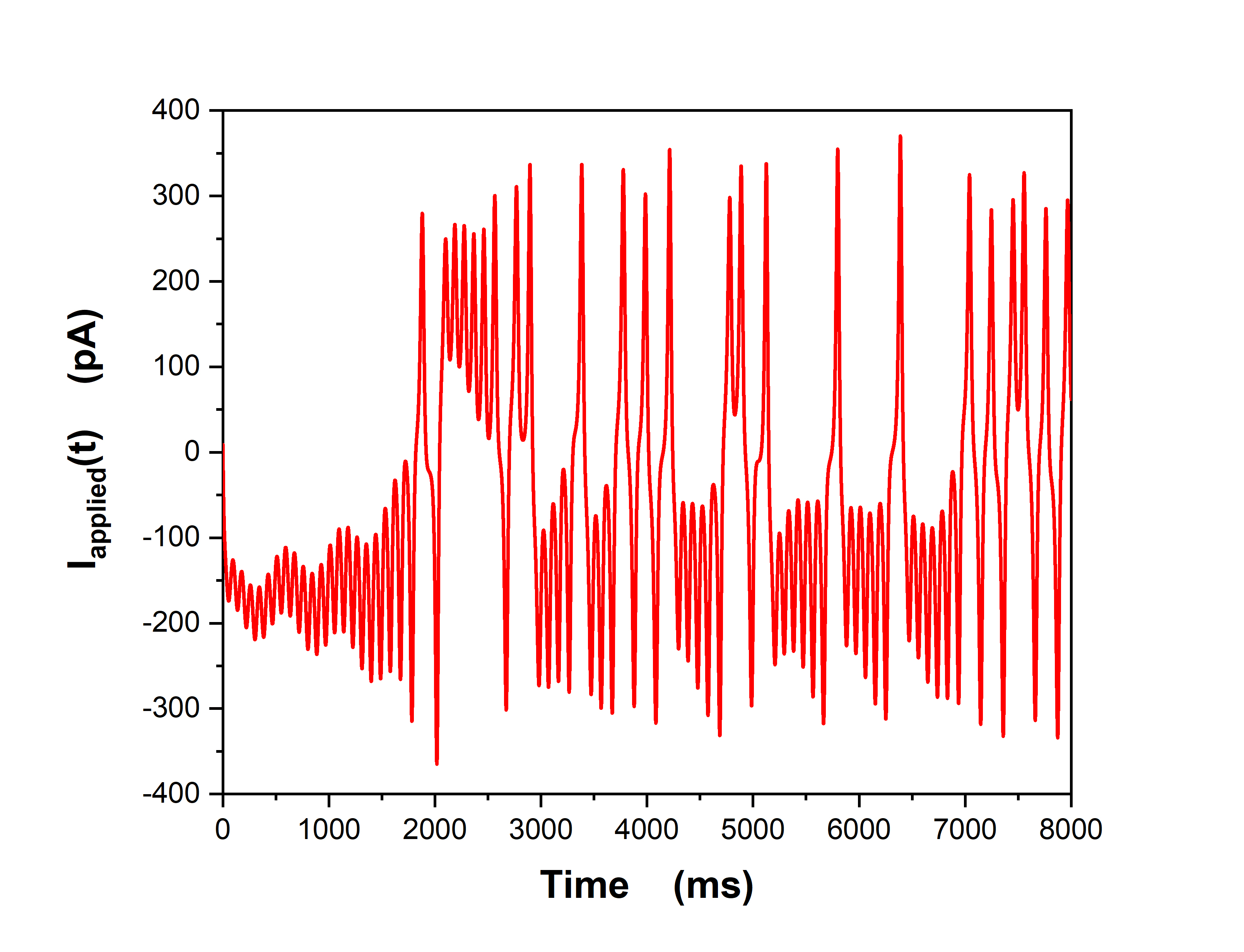}
  \caption{Stimulating current $I_{injected}(t)$ presented to the HVC$_X$ model.}
  \label{iappliedhvcx}
\end{figure}

\begin{figure}[tbph] 
  \centering
  \includegraphics[bb=0 0 3216 2461,width=5.67in,height=4.34in,keepaspectratio]{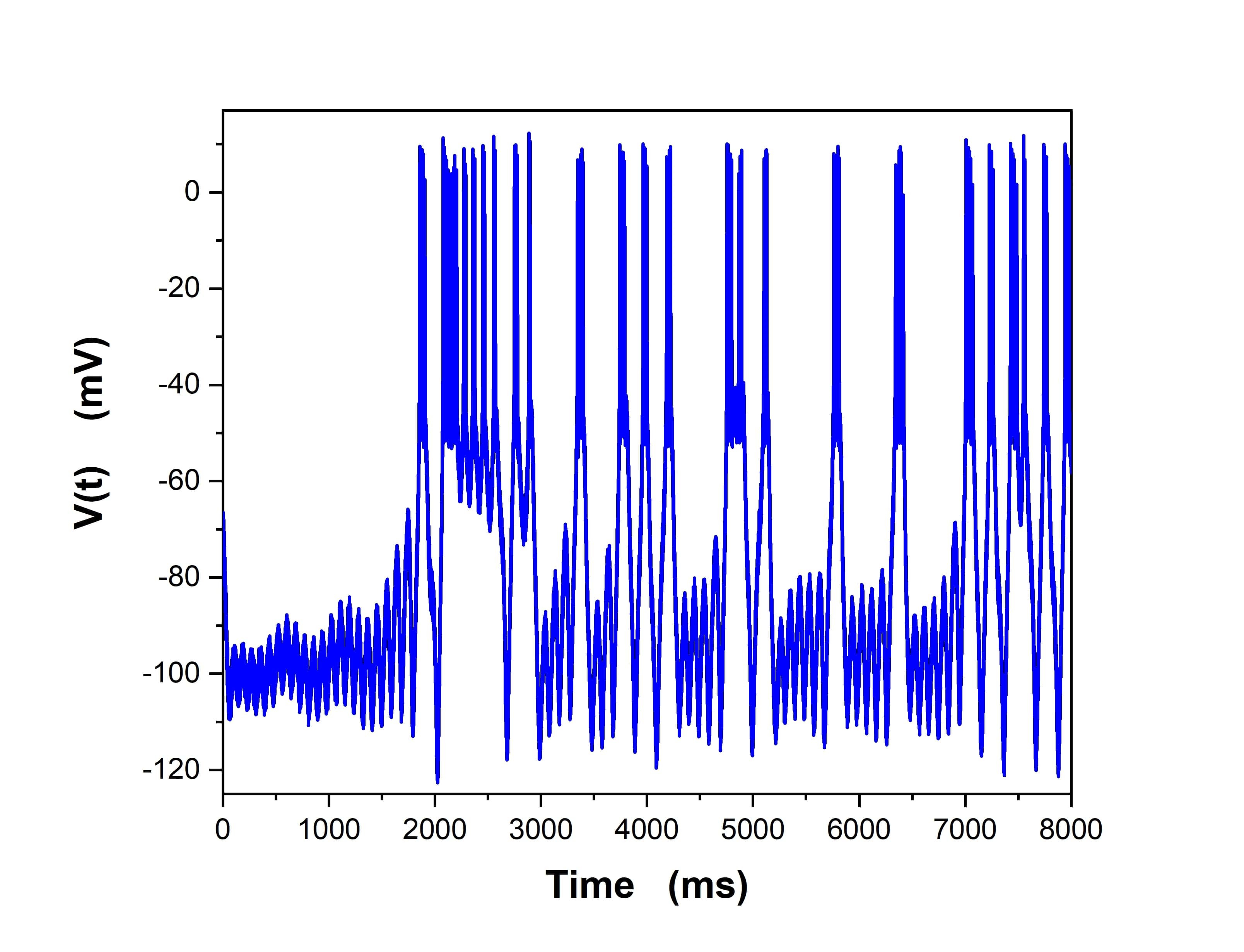}
  \caption{Response of the HVC$_X$ model membrane voltage to the selected $I_{injected}(t)$. The displayed time course $V(t)$ has no added noise.}
  \label{voltshvcx}
\end{figure}

\begin{figure}[htbp] 
  \centering
  \includegraphics[bb=0 0 3216 2461,width=5.67in,height=4.34in,keepaspectratio]{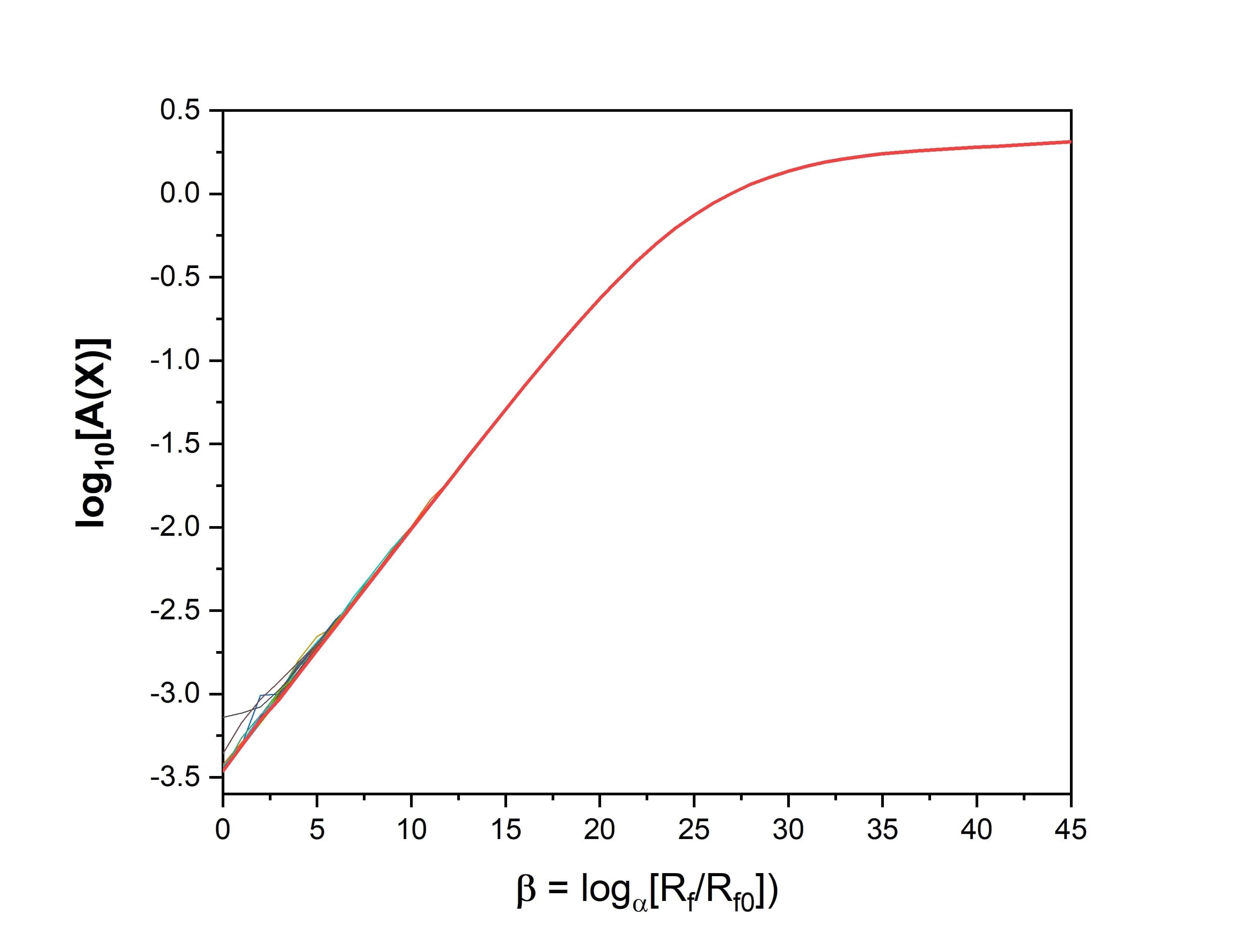}
  \caption{Action Levels of the standard model action for the HVC$_X$ neuron model discussed in the text. We see that the action rises to a `plateau' where it becomes quite independent of $R_f$. The calculation of the action uses PA with $\alpha = 1.4$ and $R_{f0} = R_m$. $N_I = 100$ initial choices for the path $\X_{init}$ were used in this calculation. For small $R_f$ one can see the slight differences in action level associated with local minima of $A(\X)$.}
  \label{actionlevelshvcx}
\end{figure}

\begin{figure}[htbp] 
  \centering
  \includegraphics[bb=0 0 3216 2461,width=5.67in,height=4.34in,keepaspectratio]{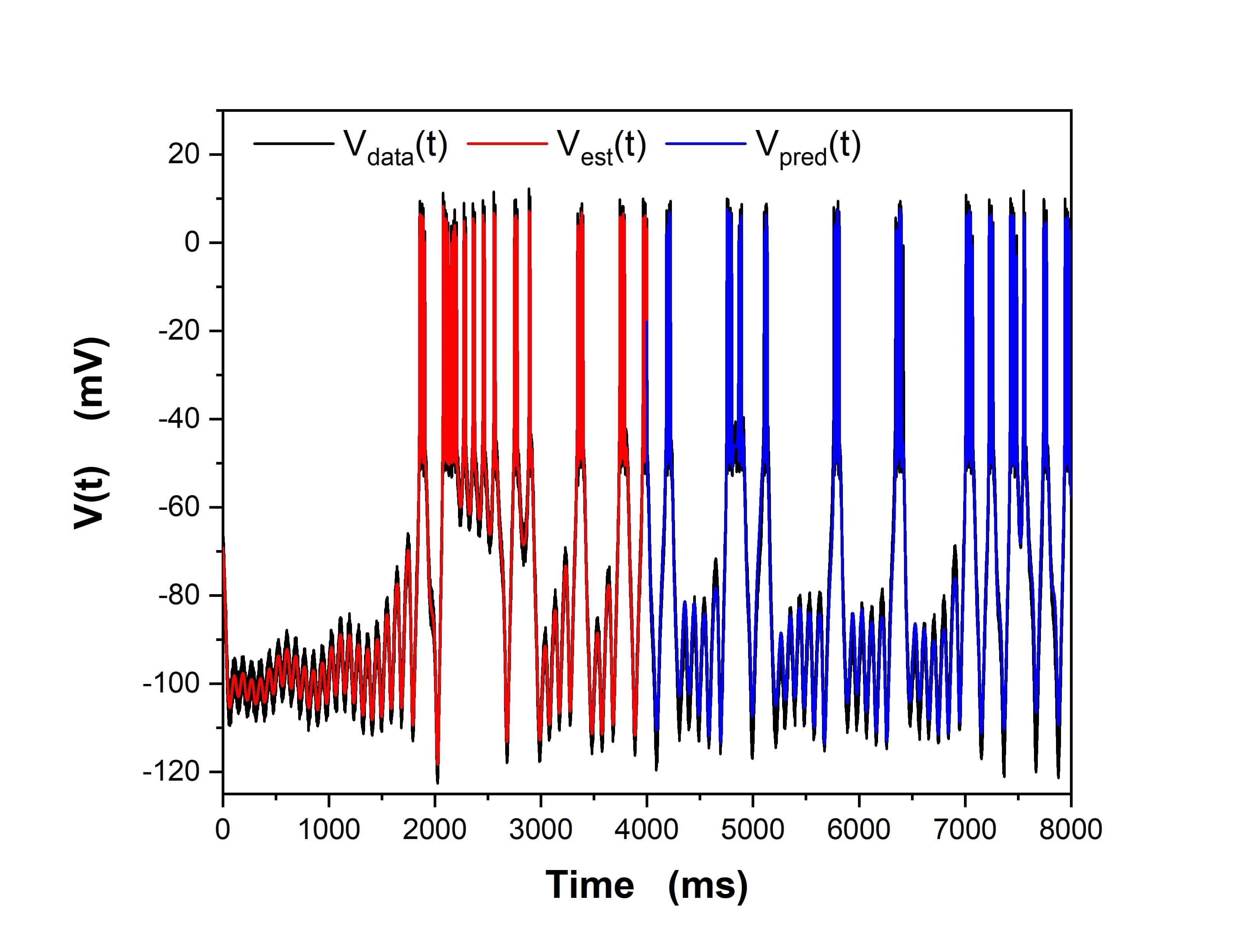}
  \caption{Results of the `twin experiment' using the model HVC$_X$ neuron described in the Appendix. Noise was added to data developed by solving the dynamical equations. The noisy $V(t)$ was presented to the precision annealing SDA calculation along with the $I_{injected}(t)$ in the observation window $t_0 = 0 \mbox{ms}, t_F = 4000 \,\mbox{ms}$. The noisy model voltage data is shown in {\bf black}, and the estimated voltage is shown in {\bf red}. For $t \ge 4000 \mbox{ms}$ we show the predicted membrane voltage, in {\bf blue}, generated by solving the HVC$_X$ model equations using the parameters estimated during SDA within the observation window.}
  \label{vdataestpredhvcx}
\end{figure}

The waveform of the injected current was chosen to have three key attributes: (1) It is strong enough to cause spiking in the neuron, (2) it dwells a long time in a hyperpolarizing region, and (3) its overall frequency content is low enough to not be filtered out by the neuron's RC time constant. On this last point, a neuron behaves like an RC circuit, it has a cut off frequency limited by the time constant of the system. Any input current which has a frequency higher than that of the cut off frequency won't be `seen' by the neuron. The time constant is taken to be the time it takes to spike and return back to 37$\%$ above its resting voltage. We chose a current where the majority of the power spectral density exists below 50 Hz.

The first two seconds of our chosen current waveform is a varying hyperpolarizing current. In order to characterize $I_h$(t) and $I_{CaT}(t)$, it is necessary to thoroughly explore the region where the current is active. $I_h(t)$ is only activated when the neuron is hyperpolarized. The activation of $I_h(t)$ deinactivates $I_{CaT}(t)$, thereby allowing its dynamics to be explored. 
In order to characterize $I_{Na}(t)$ and $I_K(t)$, it is necessary to cause spiking in the neuron. The depolarizing current must be strong enough to hit the threshold potential for spike activation. 

The parameters used to generate the data used in the twin experiment are in Table (\ref{tab:paramgen}), and the injection current data and the membrane voltage response may be seen in figures Fig. (\ref{iappliedhvcx}) and Fig. (\ref{voltshvcx}).

The numbers chosen for the data assimilation procedure in this paper are $\alpha = 1.4$ and $\beta$ ranging from 1 to 100. $R_{f,0,V} = 10^{-4}$ for voltage and $R_{f,0,j} = 1$ for all gating variables. These numbers are chosen because the voltage range is 100 times large than the gating variable range. Choosing a single $R_{f,0}$ value would result in the gating variable equations being less enforced than the voltage equation by a factor of $10^4$. The $\alpha$ and $\beta$ numbers are chosen because we seek to make $\frac{R_f}{R_{f0}}$ sufficiently large. The $\alpha$ and $\beta$ values chosen allow $\frac{R_f}{R_{f0}}$ to reach $10^{15}$.

During estimation we instructed our methods to estimate the inverse capacitance and estimate the ratio $g' = \frac{g}{C_m}$ instead of $g$ and $C_m$ independently. That separation can present a challenge to numerical procedures. We also estimated the reversal potentials as a check on the SDA method.

Within our computational capability we can reasonably perform estimates on 50,000 data points. This captures a second of data when $\Delta t = 0.02\,$ms. However, there are time constants in the model neuron which are on order 1 second. In order for us to estimate the behavior of these parameters accurately, we need to see multiple instances of the full response. We need a window on the order of 2-3 seconds. We can obtain this by downsampling the data. We know from previous results that downsampling can lead to better estimations~\citep{breen}. We take every i$^{th}$ data point, depending on the level of down sampling we want to do. In this data assimilation run, we downsampled by a factor of 4 to incorporate 4 seconds of data in the estimation window.

We look at a plot of the action as a function of $\beta$; that is, $\log[R_f/R_{f0}]$. We expect to see a leveling of of the action~\citep{ye} as a function of $R_f$. If the action becomes independent of $R_f$, we can then explore how well our parameter estimations perform when integrating them forward as predictions of the calibrated model. Looking at the action plot in Fig. (\ref{actionlevelshvcx}), we can see there is a region in which the action appears to level off, around $\beta = 40$. It is in this region where we look for our parameter estimates.

We examine all solutions around this region of $\beta$ and utilize their parameter estimates in our predictions. We compare our numerical prediction to the ``real" data from our synthetic experiment. We evaluate good predictions by finding the correlation coefficient between these two curves. This metric is chosen instead of a simple root mean square error because slight variations in spike timings yield a high amount of error even if the general spiking pattern is correct. The prediction plot and parameters for the best prediction can be seen in Fig. (\ref{vdataestpredhvcx}) and Table (\ref{tab:params}). The voltage trace in red is the estimated voltage after data assimilation is completed. It is overlayed on the synthetic input data in black. The blue time course is a prediction, starting at the last time point of the red estimated $V(t)$ trace and using the parameter estimates for $t \le 4000\,$ms.

The red curve matches the computed voltage trace quite well. There is no deviation in the frequency of spikes, spike amplitudes, or the hyperpolarized region of the cell. Looking at the prediction window, we can see that  there is some deviation in the hyperpolarized voltage trace around 9000 ms. This is an indication that our parameter estimates for currents activated in this region are not correct. Comparing parameters, we can see  that $E_h$ is estimated as lower than its actual value. Despite that, we still are able to reproduce neuron behavior fairly well.

\subsubsection{Analysis of Biophysical Parameters from HVC$_X$ Neurons in two Zebra Finch}

Our next use of SDA employs the `nudging' method described in Eq. (\ref{nudging}). In this section we used some of the data~\citep{margdaou} taken in experiments on multiple HVC$_X$ neurons from different zebra finches. The questions we asked was whether we could, using SDA, identify differences in biophysical characteristics of the birds. This question is motivated by prior biological observations~\citep{margdaou}.

Using the same HVC$_X$ model as before, we estimated the maximal conductances $\{g_{Na},g_K,g_{CaT},g_{SK},g_h\}$ holding fixed other kinetic parameters and the Nernst/Reversal potentials. The baseline characteristics of an ion channel are set by the properties of the cell membrane and the complex proteins penetrating the membrane forming the physical channel. Differences among birds would then come from the density or numbers of various channels as characterized by the maximal conductances. If such differences were identified, it would promote further investigation of the biologically exciting proposition that these differences arise in relation to some aspect of the song learning experience of the birds~\citep{margdaou}.

\begin{figure}[htbp] 
  \centering
 \includegraphics[width=5.67in,height=4.34in,keepaspectratio]{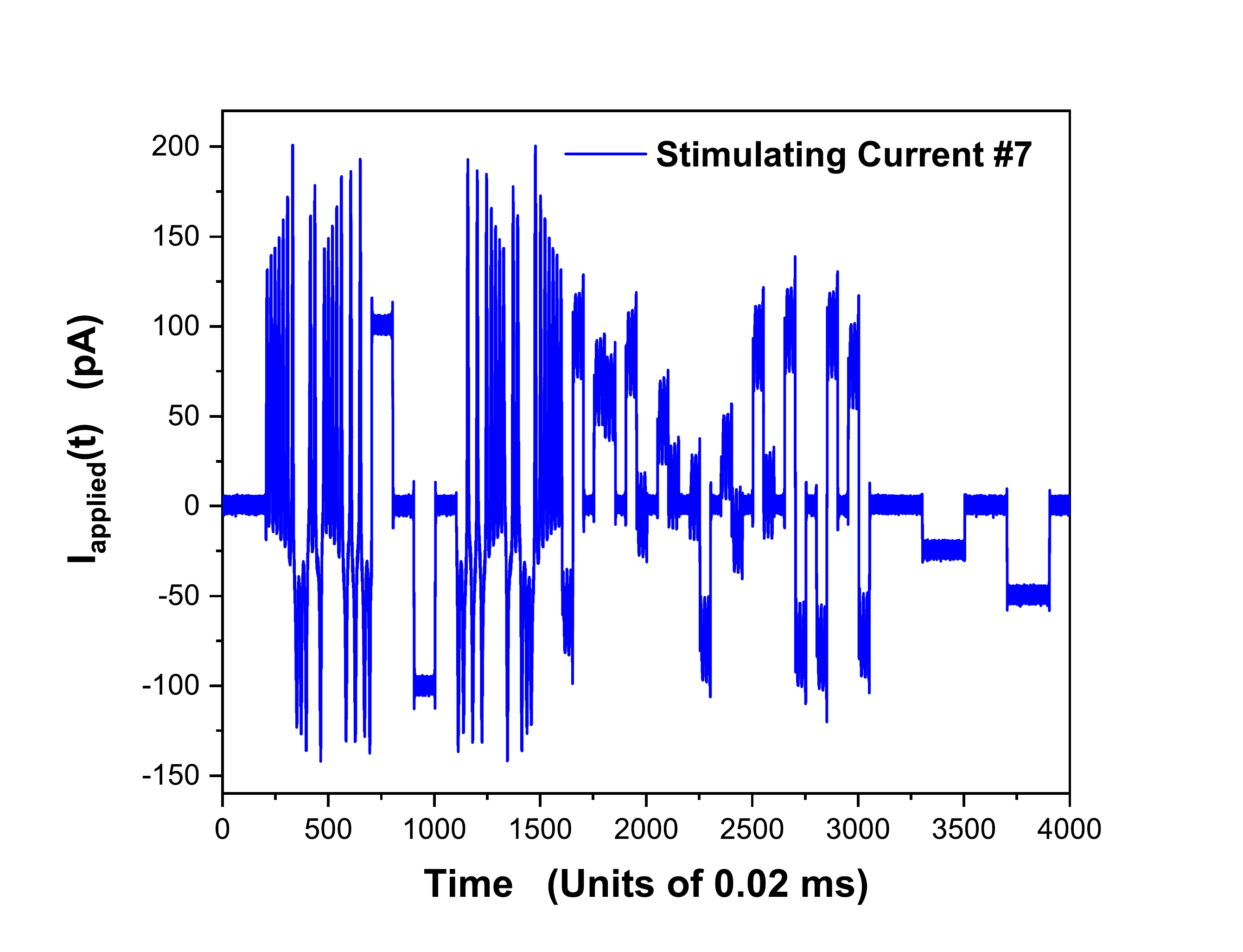}
  \caption{One of the library of stimuli used in exciting voltage response activity in an HVC$_X$ neuron. The cell was prepared {\em in vitro}, and a single patch clamp electrode injected $I_{injected}(t)$ (this waveform) and recorded the membrane potential.}
  \label{Itstimuls11hvcx}
\end{figure}

\begin{figure}[htbp] 
  \centering
  \includegraphics[width=5.67in,height=4.34in,keepaspectratio]{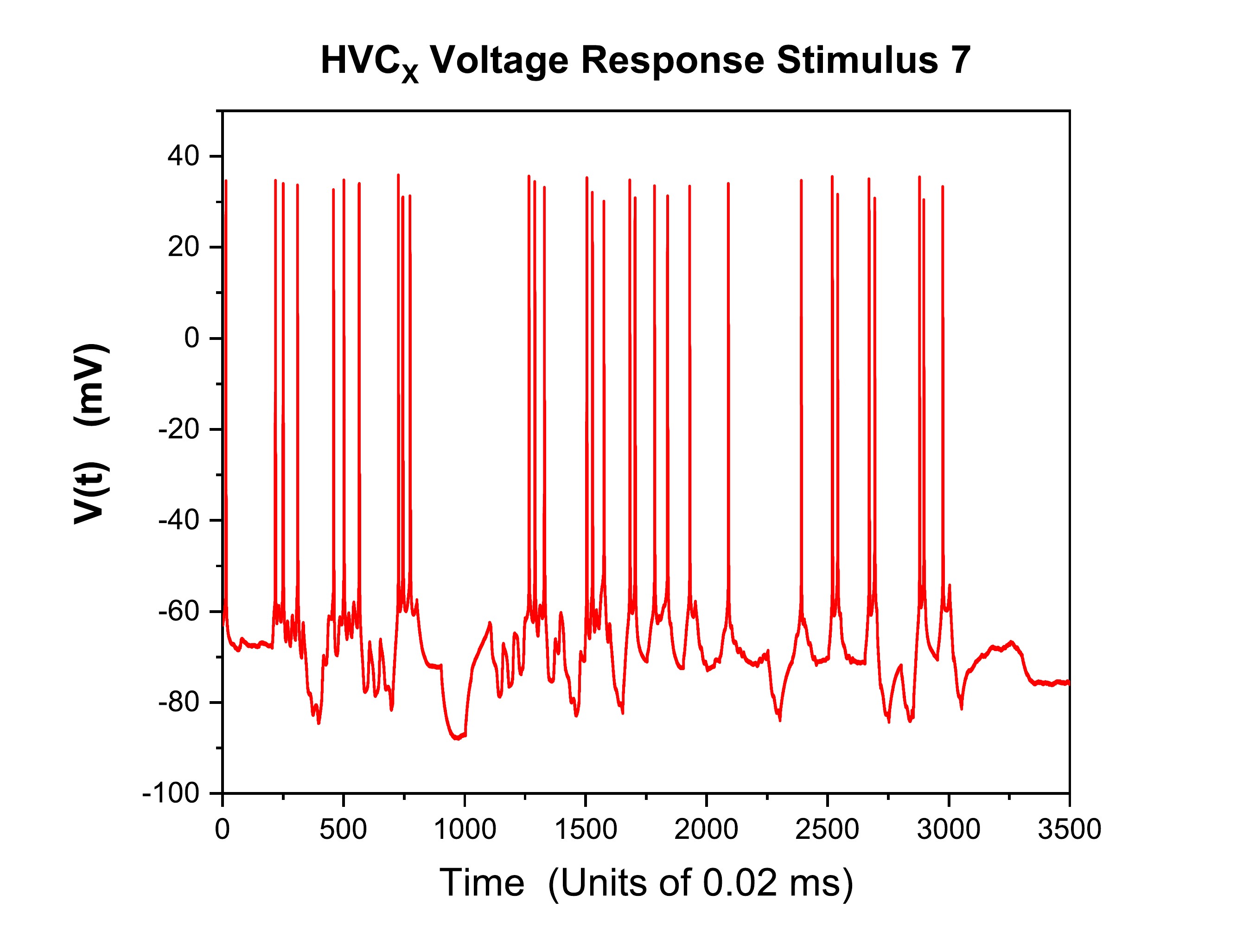}
  \caption{The voltage response from $I_{\mbox{applied}}$, Fig. (\ref{Itstimuls11hvcx}). One of the library of stimuli used in exciting voltage response activity in an HVC$_X$ neuron. The cell was prepared {\em in vitro}, and a single patch clamp electrode injected $I_{injected}(t)$ (this waveform) and recorded the membrane potential.}
  \label{Vtstimuls11hvcx}
\end{figure}

\begin{figure}[htbp] 
  \centering
  \includegraphics[width=5.67in,height=4.34in,keepaspectratio]{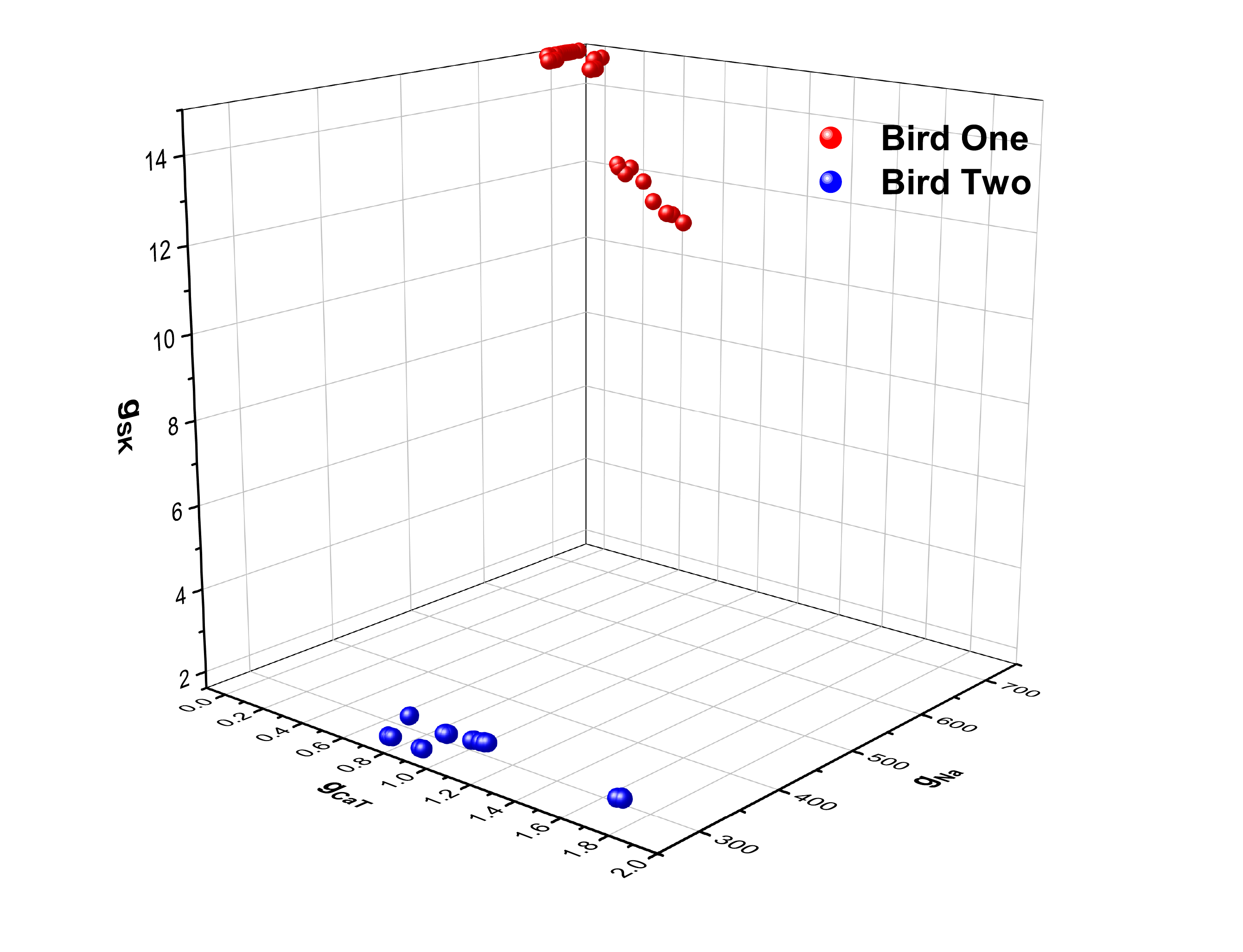}
  \caption{A three dimensional plot of three of the maximal conductances estimated from HVC$_X$ cells using the stimulating current shown in Fig. (\ref{Itstimuls11hvcx}). Membrane voltage responses from five neurons from one bird were recorded many times, and membrane voltage responses from four neurons from a second bird were recorded many times. One set of maximal conductances $\{g_{Na},g_{CaT},g_{SK}\}$ are shown. The estimates from Bird 1 are in red-like colors, and the estimates from Bird 2 are in blue-like colors. This is just one out of a large number of examples discussed in detail in~\cite{margdaou}.}
  \label{hvcxstim11gkgnagskB1red_B2blue}
\end{figure}

In Fig. (\ref{Itstimuls11hvcx}) and Fig. (\ref{Vtstimuls11hvcx}) we display the stimulating current and membrane voltage response from one of 9 neurons in our large sample. The analysis using SDA was of four neurons from one bird and seven neurons from another. The results for $\{g_{CaT},g_{Na},g_{SK}\}$ is displayed in Fig. (\ref{hvcxstim11gkgnagskB1red_B2blue}). The maximal conductances from one bird are shown in blue and from the other bird, in red. There is a striking difference between the distributions of maximal conductances.

We do not propose here to delve into the biological implications of these results. Nevertheless, we note that the neurons from each bird occupy a small but distinct region of the parameter space (Fig. \ref{hvcxstim11gkgnagskB1red_B2blue}). This result and its implications for birdsong learning, and more broadly for neuroscience, are described in~\citep{margdaou}. Here, however, we display this result as an example of the power of SDA to address a biologically important question in a systematic, principled manner beyond what is normally achieved in analyses of such data.

To fully embrace the utility of SDA for these experiments, however, further work is needed. A limitation of the present result is that the SDA estimates for $g_{SK}$ for a subset of the neurons/observations for Bird One reach the bounds of the observation window (Fig. \ref{hvcxstim11gkgnagskB1red_B2blue}). Addressing such issues would be prelude to the exciting possibility of estimating more parameters than just the principle ion currents in the Hodgkin-Huxley equations. This could use SDA numerical techniques to calculate, over hours or days, estimates of parameters that could require months or years of work to measure with traditional biological and biophysical approaches, in some cases requiring specialized equipment beyond that available for most \textit{in vitro} recording set ups. In contrast, applying SDA to such data sets requires only a computer.

\subsection{Analysis of Neuromorphic VLSI Instantiations of Neurons}

An ambitious effort in neuroscience is the creation of low power consumption analog neural-emulating VLSI circuitry. The goals for this effort range from the challenge itself to the development of fast, reconfigurable circuitry on which to incorporate information revealed in biological experiments for use in
\begin{itemize}
  \item creating model neural circuits of `healthy' performance to be compared to subsequent observations on the same circuitry informed by `unhealthy' performances. If the comparison can be made rapidly and accurately, the actual instantiations in the VLSI circuitry could be used to diagnose the changes in neuron properties and circuit connectivity perhaps leading to directions for cures, and
  \item in creating VLSI realizations of neural circuits with desired functions--say, learning syntax in interesting sequences--might allow those functions to be performed at many times increased speed than seen in the biological manifestation. If those functionalities are of engineering utility, the speed up could be critical in applications.
  \end{itemize}

One of the curious roadblocks in achieving critical steps of these goals is that after the circuitry is designed and manufactured into VLSI chips, what comes back from a fabrication plant is not precisely what we designed. This is due to the realities of the manufacturing processes and not inadequacies of the designers.

To overcome this barrier in using the VLSI devices in networks, we need an algorithmic tool to determine just what did return from the factory, so we know how the nodes of a silicon network are constituted. As each chip is an electronic device built on a model design, and the flaws in manufacuring are imperfections in the realization of design parameters, we can use data from the actual chip and SDA to estimate the actual parameters on the chip.

SDA has an advantageous position here. If we present to the chip input signals with much the same design as we prepared for the neruobiological experimets discussed in the previous section, we can measure everything about each output from the chip and use SDA to estimate the actual parameters produced in manufacturing. Of course, we do not know those paramters {\em a priori} so after estimating the parameters, thus `calibrating' the chip, we must use those estimated parameters to predict the response of the chip to a new stimulating currents. That will validate (or not) the completion of the model of the actual circuitry on the chip and permit confidence in using it in building interesting networks.

We have done this on chips produced in the laboratory of Gert Cauwenberghs at UCSD using PA~\cite{VLSI,breen} and using `nudging' as we now report.

The chip we worked with was designed to produce the simplest spiking neuron, namely one having just Na, K, and leak channels~\cite{jwu,willshaw} as in the original HH experiments. This neuron has  four state variables $\{V(t),m(t),h(t),n(t)\}$:
\bea 
&&C\frac{dV(t)}{dt} = g_{Na}m^3(t)h(t)(E_{Na} - V(t)) + g_Kn^4(t)(E_K - V(t)) \nonumber \\
&&+ g_L(E_L - V(t)) + I_{injected}(t) \nonumber \\
&&\mbox{in which the gating variables $w(t) =\{m(t),h(t),n(t)\}$ satisfy}\nonumber \\
&&\frac{d w(t)}{dt} = \frac{(w_{\infty}(V(t)) - w(t))}{\tau_w(V(t))},
\eea
and the functions $w_{\infty}(V)$ are discussed in depth in~\cite{jwu,willshaw}.

\begin{figure}[htbp] 
  \centering
  \includegraphics[bb=0 0 3216 2461,width=5.67in,height=4.34in,keepaspectratio]{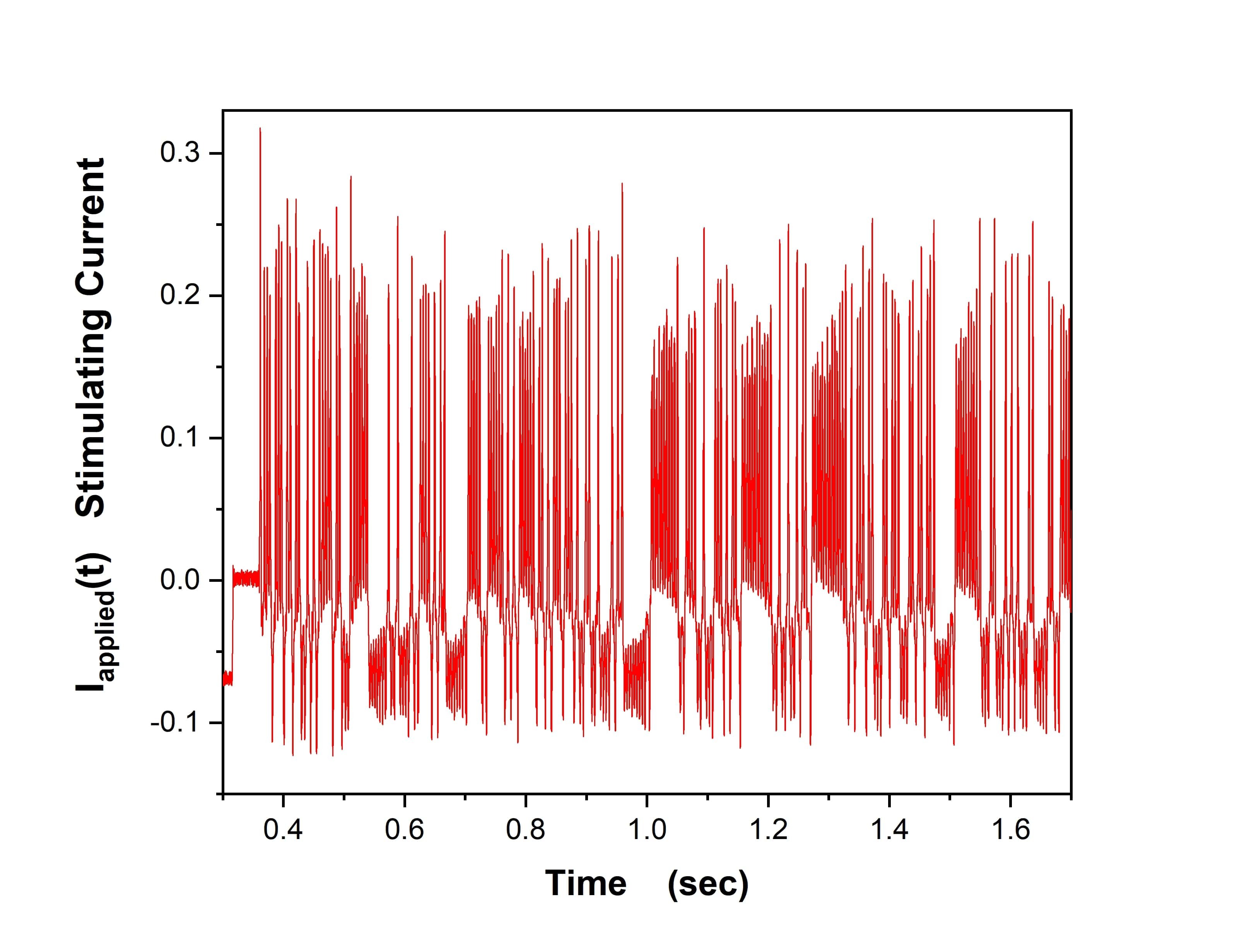}
  \caption{This waveform for $I_{injected}(t)$ was used to drive the VLSI NaKL neuron after receipt from the fabrication facility.}
  \label{vlsi_nakl_iapplied}
\end{figure}

In our experiments on a `NaKL' chip we used the stimulating current displayed in Fig. (\ref{vlsi_nakl_iapplied}),

and measured {\bf all} the neural responses $\{V(t),m(t),h(t),n(t)\}$. These observations were presented to the designed model within SDA to estimate the parameters in the model.

\begin{figure}[htbp] 
  \centering
   \includegraphics[bb=0 0 3216 2461,width=5.67in,height=4.34in,keepaspectratio]{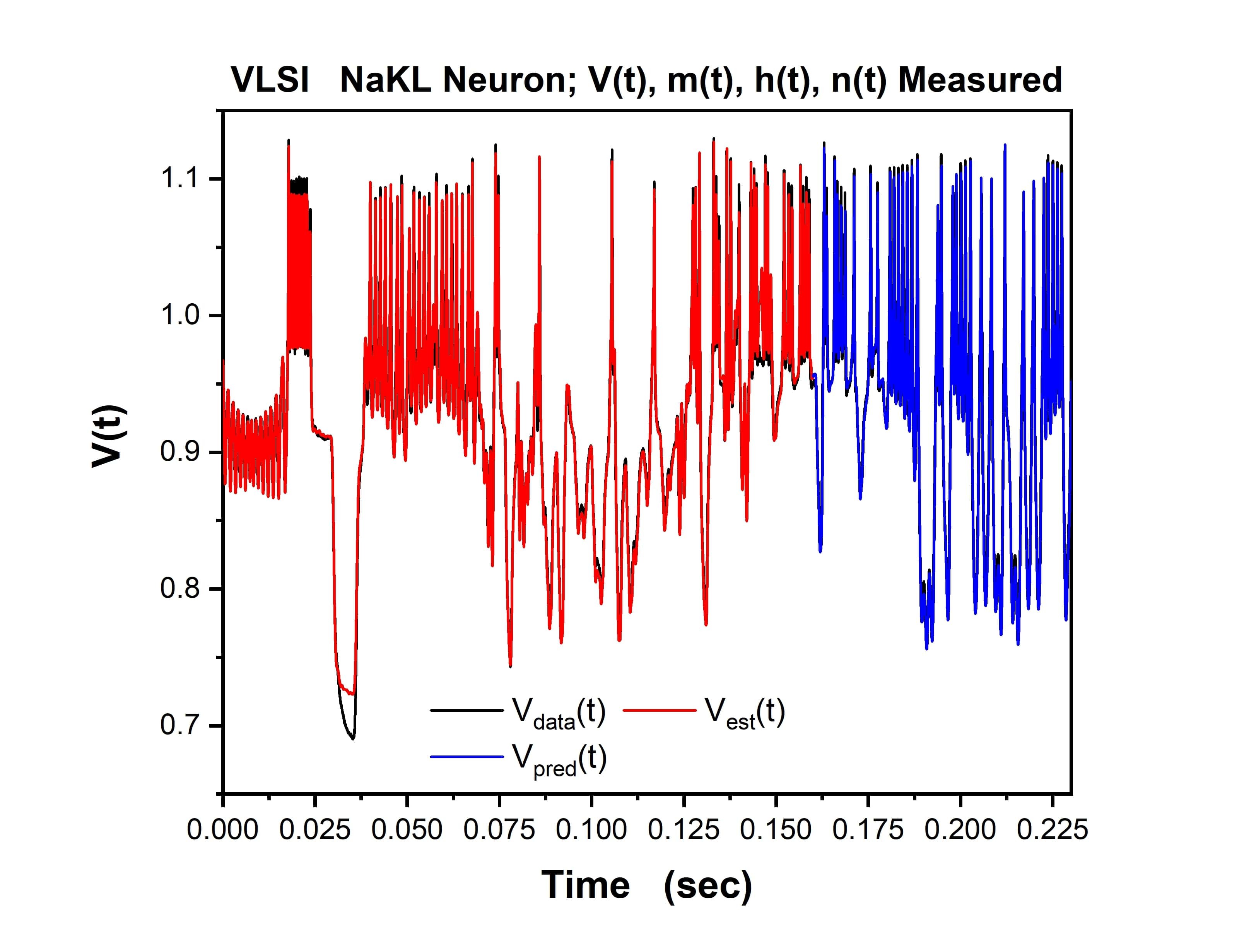}
  \caption{The NaKL VLSI neuron was driven by the waveform for $I_{injected}(t)$ seen in Fig. (\ref{vlsi_nakl_iapplied}). The four state variables $\{V(t),m(t), h(t),n(t)\}$ for the NaKL model were recorded and used in an SDA `nudging' protocol to estimate the parameters of the model actually realized at the manufacturing facility. Here we display the membrane voltages: $\{V_{data}(t), V_{est}(t),V_{pred}(t)\}$--the observed membrane voltage response when $I_{injected}(t)$ was used, the estimated voltage response using SDA, and finally the predicted voltage response $V_{pred}(t)$ from the calibrated model actually on the VLSI chip. In a laboratory experiment, only this attribute of a neuron would be {\bf observable}.}
  \label{vlsi_naklv_data_est_pred}
\end{figure}

We then tested/validated the estimations by using the calibrated model to predict how the VLSI chip would respond to a different injected current. In Fig. (\ref{vlsi_naklv_data_est_pred}) we show the observed $V_{data}(t)$ in black, the estimation of the voltage through SDA in red, and the prediction of $V(t)$ in blue for times after the end of the observation window. 

While one can be pleased with the outcome of these procedures, for our purposes we see that the use of our SDA algorithms gives the user substantial confidence in the functioning characteristics of the VLSI chips one wishes to use at the nodes of a large, perhaps even very large, realization of a desired neural circuit in VLSI. We are not unaware of the software developments to allow efficient calibration of very large numbers of manufactured silicon neurons. A possible worry about also determining the connectivity, both the links and their strength and time constants, may be alleviated by realizing these links through a high bandwidth bi-directional connection  of the massive array of chips and the designation of connectivity characteristics on an off-chip computer.

\begin{figure}[htbp] 
  \centering
    \includegraphics[bb=0 0 3216 2461,width=5.67in,height=4.34in,keepaspectratio]{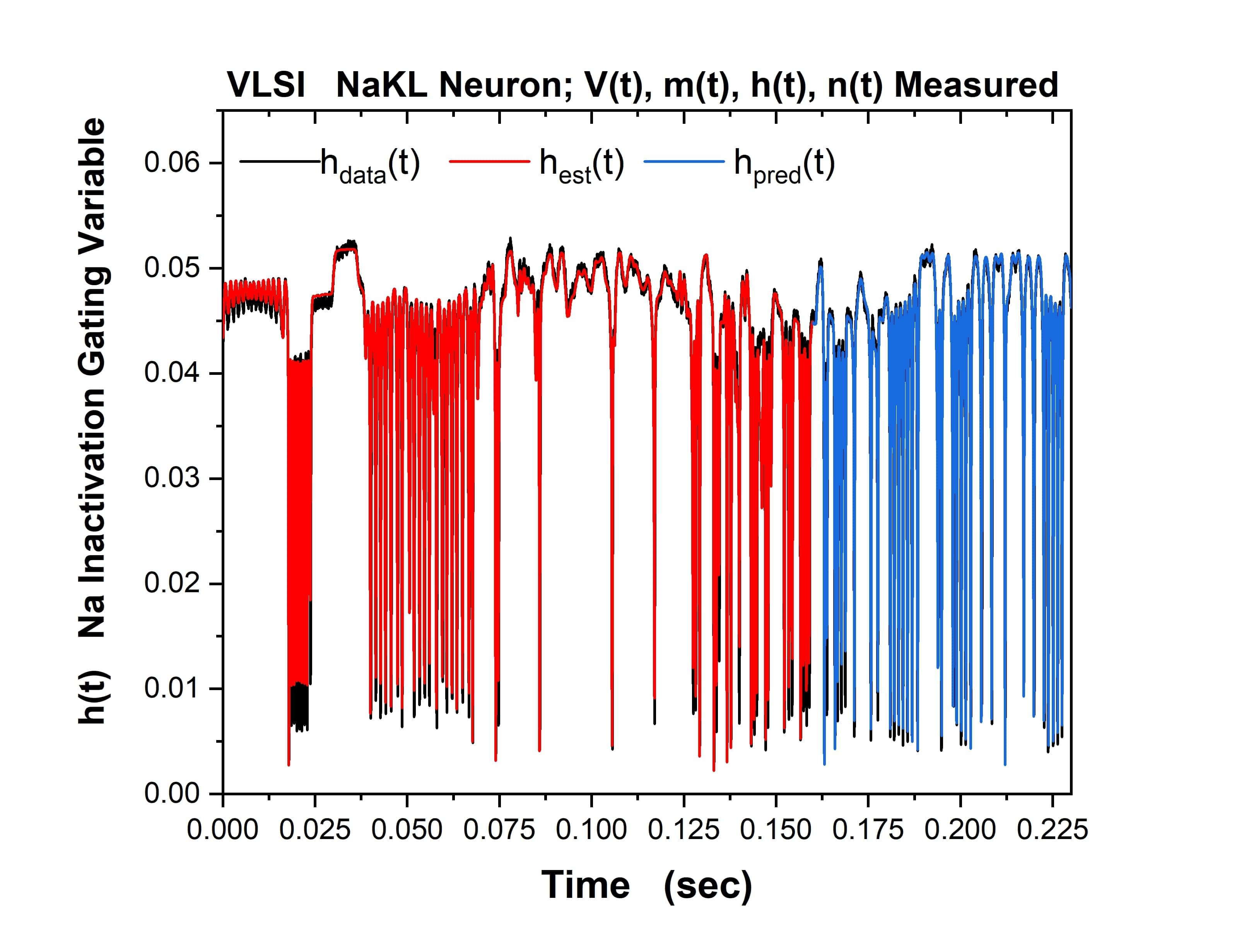}
  \caption{The NaKL VLSI neuron was driven by the waveform for $I_{injected}(t)$ seen in Fig. (\ref{vlsi_nakl_iapplied}. The four state variables $\{V(t),m(t), h(t),n(t)\}$ for the NaKL model were recorded and used in an SDA `nudging' protocol to estimate the parameters of the model actually realized at the manufacturing facility. Here we display the Na inactivation variable $h(t)$: $\{h_{data}(t), h_{est}(t),h_{pred}(t)\}$--the observed $h(t)$ time course when $I_{injected}(t)$ was used, the estimated $h(t)$ time course  using SDA, and finally the predicted $h(t)$ time course from the calibrated model actually on the VLSI chip. In a laboratory experiment, this attribute of a neuron would be {\bf unobservable}. Note we have rescaled the gating variable from its natural range $0 \le h(t) \le 1)$ to the range within the VLSI chip. The message of this Figure is in the very good accuracy and prediction of an experimentally unobservable time course.}
  \label{vlsi_nakl_hdata_est_pred}
\end{figure}

Part of the same analysis is the ability to observe, estimate and predict the experimentally unobservable gating variables. This serves, in this context, as a check on the SDA calculations. The Na inactivation variable $h(t)$ is shown in Fig. (\ref{vlsi_nakl_hdata_est_pred}) as its measured time course $h_{data}(t)$ in black, its estimated time course $h_{est}(t)$ in red, and its predicted time course $h_{pred}(t)$ in blue.

\section{Discussion}

Our review of the general formulation of statistical data assimilation (SDA) started our remarks. Many details can be found in~\cite{abar13} and subsequent papers by the authors. Recognizing that the core problem is to perform, approximately of course, the integral in Eq. (\ref{expect}) is the essential take away message. This task requires well `curated' data and a model of the processes producing the data. In the context of experiments in life sciences or, say, aquisition of data from earth system sensors, curation includes an assessment of errors and the properties of the instruments as well. 

One we have the data and a model, we still need a set of procedures to transfer the information from the data to the model, then test/validate the model on data not used to train the model. The techniques we covered are general. Their application to examples from neuroscience comprise the second part of this paper.

In the second part we first address properties of the avian songbird song production pathway and a neural control pathway modulating the learning and production of functional vocalization--song. We focus our attention on one class of neurons, HVC$_X$, but have also demonstrated the utility of SDA to describe the response properties of other classes of neurons in HVC, such as HVC$_{RA}$~\cite{Nirag2016} and HVC$_I$~\cite{Breen2016HVCI}. Indeed, the SDA approach is generally applicable wherever there is insight to relate biophysical properties of neurons to their dynamics through Hodgkin Huxely equations.

Our SDA methods considered variational algorithms that seek the highest conditional probability distributions of the model states and parameters conditioned on the collection of observations over a measurement window in time. Other approaches, especially Monte Carlo algorithms were not discussed here, but are equally attractive.

We discussed methods of testing models of HVC$_X$ neurons using ``twin experiments" in which a model of the individual neuron produces synthetic data to which we add noise with a selected signal-to-noise-ratio. Some state variable time courses from the library of these model produced data, for us the voltage across the cell membrane, is then part of the action Eq. (\ref{standard}), specifically in the measurement error term. Errors in the model are represented in the model error term of the action.

Using a precision annealing protocol to identify and track the global minimum of the action, the successful twin experiment gives us confidence in this SDA method from information trans from data to the model. 

We then introduced a `nudging' method as an approximation to the Euler-Lagrange equations derived from the numerical optimization of the action Eq. (\ref{standard})--this is Laplace's method in our SDA context. The nudging method, introduced in meteorology some time ago, was used to distinguish between two different members of the Zebra Finch collection. We showed, in a quite preliminary manner, that the two, unrelated birds of the same species, express different HVC network properties as seen in a critical set of maximal conductances for the ion channels in their dynamics.

Finally we turned to a consideration of the challenge of implementing in VLSI technology the neurons in HVC towards the goal of building a silicon-HVC network. The challenge at the design and fabrication stages of this effort where illuminated by our use of SDA to determine what was actually returned from the manufacturing process for our analog neurons.

\subsection{Moving Forward to Network Analysis}

Finally, we have a few comments associated with the next stage of analysis of HVC. In this, and previous papers, we analyzed individual neurons in HVC. These analyses were assisted by our using SDA, through twin experiments, to design laboratory experiments though the selection of effective stimulilaing injected currents.

\begin{figure}[tbhp] 
  \centering
  \includegraphics[keepaspectratio]{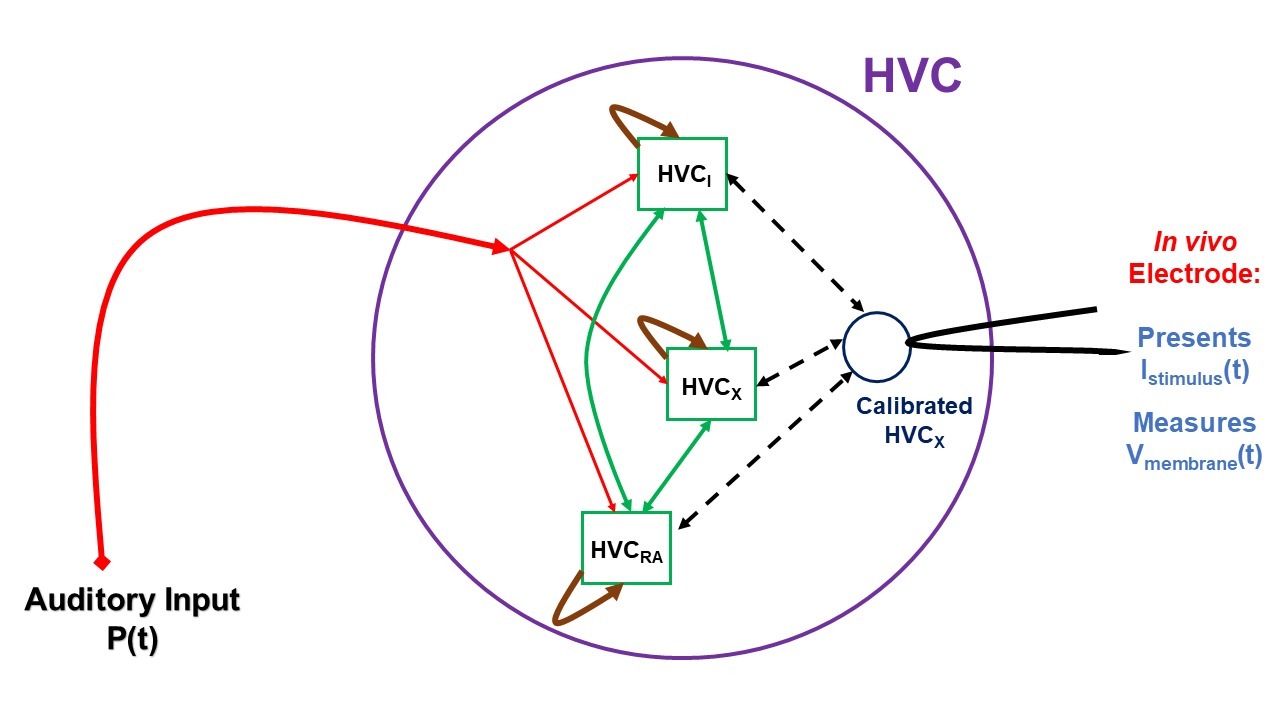}
  \caption{A cartoon-like idea of an experiment to probe the HVC network. In this graphic three neuron populations of $\{HVC_X,HVC_I,HVC_{RA}\}$ neurons are stimulated by auditory signals $P(t)$ presented to the bird {\em in vivo}. This drives the auditory pathway from the ear to HVC, and the network activity is recorded from a calibrated, living HVC$_X$ neuron, used here as a sensor for network activity. While the experiment is now possible, the construction of a model HVC network will proceed in steps of complexity using simple then more biophysiclly realistic neuron models and connections among the nodes (neurons) of the network. From libraries of time courses of $P(t)$, chosen by the user, and responses of $V(t)$ in the `sensor' HVC$_X$ neuron, we will use SDA to estimate properties of the network.}
  \label{HVC_graphic_05_05_18}
\end{figure}

Having characterized the electrophysiology of an individual neuron within the framework of Hodgkin-Huxley (HH) models, we may now proceed beyond the study of individual neurons~\cite{scirept2016} {\em in vitro}. Once we have characterized an HVC neuron through a biophysical HH model, we may then use it {\em in vivo} as a sensor of the activity of the HVC network where it is connected to HVC$_{RA}$, HVC$_I$, and other HVC$_X$ neurons. The schema for this kind of experiment is displayed in Fig. (\ref{HVC_graphic_05_05_18}). These experiments require the capability to perform measurements on HVC neurons in the living bird. That capability is available, and experiments as suggested in our graphic are feasible, if challenging.

The schematic indicates that the stimulating input to the experiments is auditory signals, chosen by the user, presented to the bird's ear and reaching HVC through the auditory pathway. The stimuli from this signal is then distributed in a manner to be deduced from experiment and then produces activity in the HVC network that we must model. The goal is, at least initially, to establish, again within the models we develop, the connectivity of HVC neuron classes as it manifests itself in the function of the network. We have some information about this~\cite{Mooney2005,Kosche15}, and these results will guide the development of the HVC model used in these whole-network experiments. An important point to address is what changes to the {\em in vitro} model might be necessary to render it a model for {\em in vivo} activity.

\newpage

\section*{Appendix}

The equations describing the HVC$_X$ neuron dynamics are taken from the work of Daou~\cite{Daou2013}. That paper also has an extensive account of his experiments on the other two major classes of neurons in HVC. The equations are of Hodgkin-Huxley (HH) form for a neuron without spatial extent; this is called a one-compartment model. It is meant to apply to neurons in isolation of the network, here HVC, in which they sit {\em in vivo}. The dynamical variables include the observable quantities: voltage across the cell membrane, $V(t)$ and the intracellular concentration of $[Ca^{2+}]_{in}(t) = C(t)$. $V(t)$ is directly connected to action potentials or voltage spikes that communicate among cells in a network; the time scale of these spikes is a few ms. $C(t)$ provides a slow background modulation that raises the cells potential (depolarizes the cell) or lowers it (hyperpolarizes the cell) on time scales as long as 10's of ms.

The voltage equation, conservation of charge, relates the capacitance of the cell membrane $C_m$ as it separates concentrations of ions within and without the cells to the various currents which contain the nonlinear voltage dependence of the permeability of ions to passing into and out of the cell. The model represents these ion currents: $\{Na, K, Ca\}$ in several different ways.

The general form of an HH current is 
\be 
I_{ion}(t) = g_{ion}m^{integer1}_{ion}(t)h^{integer2}_{ion}(t)(E_{rev-ion} - V(t)),
\ee
where the reversal potential is the equilibrium Nernst potential~\cite{jwu,willshaw}. The gating variables  $\{m(t),h(t)\}$ lie between zero and one and represent the amount the ion channel is open relative to the maximum opening it may have. The maximal conductance $g_{ion}$ represent the number or density of ion channels in the neuron model. This form of ion current applies when the concentrations of the ion are not significantly different outside and inside the cell. This is not so for $Ca^{2+}$ ions, so there we use the Goldman-Hodgkin-Katz (GHK) form of the current~\cite{Goldman43}.
\be
I_a(t) = -P_a z^2_a F^2 \left(\frac{[ion]_{in}(t) - [ion]_{out}e^{-z_aFV(t)/RT}}{1-e^{-z_aFV(t)/RT}}\right),
\ee
for ion$a$. $z_a$ is the charge on the ion, $F$ the Faraday constant, $R$ the gas constant, $T$ the temperature, and $P_a$ the permeability of the cell membrane to ion $a$.

We use an approximation to the GHK equations for the two types of Calcium currents selected in~\cite{Daou2013}
\bea
I_{CaL}(t) &=& g_{CaL}Vs_{\infty}^2(V(t)) \left( \frac{[Ca]_{out}}{e^{2FV(t)/RT}-1} \right) \nonumber \\
I_{CaT}(t) &=& g_{CaT}V(t)[a_T]_{\infty}^3(V)[b_T]_{\infty}^3(r_T^A) \left(\frac{[Ca]_{out}}{e^{2FV(t)/RT}-1}\right) \\ \nonumber
&&b_{T_{\infty}}(r_T) = \frac{1}{1+e^{\left(\frac{r_T - \theta_b}{\sigma_b}\right)}} - \frac{1}{1+e^{\left(\frac{-\theta_b}{\sigma_b}\right)}}
\eea
$a_T$ and $b_T$ are instantaneous activating and inactivating gating variables, respectively. $r_T$ is a slow gating variable which takes the same functional form as $a_T$ and other gating variables $m(t)$ and $h(t)$. These gating variables $w(t)$ satisfy a first order kinetic equation
\be
\frac{d w(t)}{dt} = \frac{w_{\infty}(V(t)) - w(t))}{\tau_w(V(t))},
\ee
in which 
\begin{equation}
w_{\infty}(V) = \frac{1}{2}\left[1-\tanh\left(\frac{V-\theta_w}{2\sigma_w}\right)\right], 
\end{equation}
for all gating variables except $h_{\infty}(V)$ appearing in $I_{Na}(t)$~\cite{Daou2013}.
$\theta_w$ is the half-activation voltage and $\sigma_w$ controls the slope of the activation function. 
For fast gating variables, such as $m$ of $I_{Na}$, and $s$ of $I_{CaL}$ we replace the time dependence by $W_{\infty}(V)$. 

$\tau_w(V)$ is the time constant of each gating variable. Time constants for the $n$ and $hp$ gating variables (these names refer to~\cite{Daou2013}) are given below, where $\bar{\tau}_w$ is an average time constant. Our model differs from~\cite{Daou2013} by one time constant. Instead, $\tau_{rs}(V)$ takes the form presented here:
\bea
&&\tau_w(V) = \frac{\bar{\tau}_w}{\cosh\left(\frac{V - \theta_w}{2\sigma_w}\right)} \nonumber \\
\mbox{ for  n  or  hp} \nonumber \\
&&\tau_{rs}(V) = 0.1+193.0\left(1-\tanh^2\left(\frac{V(t)+80}{-21}\right)\right) \nonumber \\
&&\tau_{rf} = \frac{p_{r_f}}{\frac{-7.4\left(V+70\right)}{e^{\frac{V+70}{-0.8}}-1}+65e^{\frac{V+56}{-23}}}\;\;\tau_{r_T}(V) = \tau_{r_0} + \frac{\tau_{r_1}}{1+e^{\left(\frac{V-\theta_{r_T}}{\sigma_{r_T}}\right)}} \nonumber
\eea
For our choice of ion currents we follow the results of experimental data~\cite{Dutar98,Mooney2000,KubotaTaniguchi98} and generally reproduce the model 
listed in~\cite{Daou2013}. HVC$_{{X}}$ spiking properties include fast rectifying current, sag in response to hyperpolarizing current, and spike frequency adaption in response to depolarizing current.
\bea
\label{eq:HVC}
&&C\frac{dV(t)}{dt} = I_{Na}(t) + I_K(t) + I_L(t) + I_{CaT}(t) + I_{CaL}(t) \nonumber \\
&&+ I_A(t) + I_{SK}(t) + I_h(t) + I_{Nap}(t) + I_{injected}(t)
\eea

$I_{Na}(t)$ and $I_{K}(t)$ are the standard HH currents. They produce fast spiking in response to injected currents. $I_L(t)$ is a leak current meant to capture all linear currents of the neuron.  $I_{CaT}(t)$ is a low threshold T-type calcium current that causes rebound depolarization in cooperation with $I_h(t)$. $I_{CaL}(t)$ is a high threshold L-type calcium current. $I_{CaL}(t)$ works in conjunction with $I_{SK}(t)$, a calcium concentration dependent potassium current, to create frequency adaptation in neuron spiking. $I_A(t)$ is an A-type potassium current. $I_{Nap}(t)$ is a persistent sodium current. From the model presented in \cite{Daou2013}, we eliminate $I_{KNa}(t)$, a sodium dependent potassium current, and rewrite all sigmoidal functions as hyperbolic tangents.

The mass conservation equation for $Ca^{2+}$ is written as
\be
\frac{d C(t)}{dt} = \epsilon(I_{CaT}(t) + I_{CaL}(t)) + k_{Ca}(b_{Ca} - C(t)),
\label{eq:dcadt}
\ee
again following~\cite{Daou2013}.

\section*{Tables}
{
\begin{table}[!h]
\centering
\begin{tabular}{|l|c|l|c|l|c|}
\hline
\rule{0pt}{2ex}
Parameter & Value & Parameter & Value& Parameter & Value \\
\hline
\rule{0pt}{2ex}
$g_{Na}$  & 450 nS & $g_L$ & 2 nS & $k_f$ & 0.3\\[1pt]
$E_{Na}$ & 50 mV & $E_L$ & -70 mV & $\theta_{mp}$ & -40 mV \\[1pt]
$g_K$    & 50 nS & $g_{Nap}$ & 1 nS & $\sigma_{mp}$ & -6 mV \\[1pt]
$E_K$ & -90 mV & $g_{CaL}$ & 19 nS & $\theta_s$ & -20 mV \\[1pt]
$g_{CaT}$ & 2.65 nS & $\theta_m$ & -35 mV & $\sigma_s$ & -0.05 mV \\[1pt]
$g_{SK}$ & 6 nS & $\sigma_m$ & -5 mV & $\theta_{hp}$ & -48 mV \\[1pt]
$g_{H}$ & 4 nS & $\theta_n$ & -30 mV & $\sigma_{hp}$ & 6 mV \\[1pt]
$E_H$ & -30 mV &$\sigma_n$ & -5 mV & $\bar{\tau}_{hp}$ & 1000 ms \\[1pt]
$C_m$  & 100 pF & $\bar{\tau}_n$ & 10 ms  & $\theta_e$ & -60 mV \\[1pt]
$\theta_a$ & -20 mV & $\theta_{rf}$ & -105 mV & $\sigma_e$ & 5 mV \\[1pt]
$\sigma_a$ & -10 mV & $\sigma_{rf}$ & 5 mV & $\tau_e$ & 20 ms \\[1pt]
$\theta_{rs}$ & -105 mV & $\theta_{aR}$ & -65 mV & $\theta_{b}$ & 0.4 mV \\[1pt]
$\sigma_{rs}$ & 25 mV & $\sigma_{aT}$ & -7.8 mV & $\sigma_b$ & -0.1 mV\\[1pt]
$\theta_{rT}$ & -67 mV & $\theta_{rrT}$ & 68 mV & $f$ & 0.1 \\[1pt]
$\sigma_{rT}$ & 2 mV & $\sigma_{rrT}$ & 2.2 mV & $\epsilon$ & 0.0015 $\frac{\mu { M}}{{pA} \cdot {ms}}$ \\[1pt]
$\tau_{r_0}$ & 200 ms & $\tau_{r_1}$ & 87.5 ms & $p_{r_f}$ & 100 \\[1pt]
$k_{Ca}$ & 0.3 ms$^{-1}$ & $b_{Ca}$ & 0.1 $\mu$M & $k_s$ & 0.5  $\mu$M \\[1pt]
\hline
\end{tabular}
\caption{\label{tab:paramgen}Parameter values used to numerically generate the HVC$_X$ data. The source of these values comes from~\citep{Daou2013}. Data was generated using an adaptive Runge-Kutta method, and can be seen in Fig. (\ref{iappliedhvcx}) and Fig. (\ref{voltshvcx}).}
\end{table}
}

{
\begin{table}[!h]
\centering
\begin{tabular}{|l| c| c| c| l|}
\hline 
\rule{0pt}{2ex}
Parameter & Bounds & Best Estimate & Actual Value & Units\\
\hline 
\rule{0pt}{2ex}
$g_{Na}'$  & 0.1, 10  & 4.98 & 4.5 & nS/pF \\[1pt]
$E_{Na}$ & 1, 100 & 43.2 & 50 & mV \\[1pt]
$g_K'$    & 0.01, 5 & 0.907 & 0.5 & nS/pF \\[1pt]
$E_K$ & -140, -10 & -127.4 & -90 & mV \\[1pt]
$g_{CaT}'$ & 0.001, 1 & 0.0326 & 0.0265 & nS/pF\\[1pt]
$g_{SK}'$ & 0.001, 1 & 0.0373 & 0.06 & nS/pF\\[1pt]
$g_{h}'$ & 0.001, 1 & 0.0432 & 0.04 & nS/pF\\[1pt]
$E_h$ & -100, -1 & -44.1 & -30 & mV \\[1pt]
$C_{inv}$  & 0.001, 0.5 & 0.011 & 0.01 & pF$^{-1}$ \\[1pt]
\hline
\end{tabular}
\caption{\label{tab:params}Parameter Estimates from the Best Predictions. The best prediction is chosen by finding the highest correlation coefficient between the predicted voltage and ``real" voltage. This comparison can be made on experimental data. It represents an attractive alternative to the familiar least squares metric commonly used. That metric is very sensitive to spike times in data with action potentials: small errors in spike times may result in large errors in a least squares metric.}
\end{table}
}

\newpage 

\bibliographystyle{frontiersinHLTH&FPHY} 

\begin{thebibliography}{51}
\expandafter\ifx\csname natexlab\endcsname\relax\def\natexlab#1{#1}\fi
\expandafter\ifx\csname urlstyle\endcsname\relax
  \expandafter\ifx\csname doi\endcsname\relax
  \def\doi#1{doi:\discretionary{}{}{}#1}\fi \else
  \expandafter\ifx\csname doi\endcsname\relax
  \def\doi{doi:\discretionary{}{}{}\begingroup \urlstyle{rm}\Url}\fi \fi
\expandafter\ifx\csname selectlanguage\endcsname\relax
  \def\selectlanguage#1{}\fi

\bibitem[{Lorenz(2006)}]{lor96}
Lorenz EN.
\newblock Predictability: A problem partly solved.
\newblock Palmer T, Hagedorn R, editors, {\em Predictability of weather and
  climate\/} (Cambridge) (2006).

\bibitem[{Evensen(2009)}]{even}
Evensen G.
\newblock {\em Data Assimilation: The Ensemble Kalman Filter\/} (Springer)
  (2009).

\bibitem[{Abarbanel(2013)}]{abar13}
Abarbanel HDI.
\newblock {\em Predicting the Future: Completing Models of Observed Complex
  Systems\/} (Springer) (2013).

\bibitem[{Tong(2011)}]{tong2011}
Tong D.
\newblock Statistical physics  (2011).
\newblock Available at {\em www.damtp.cam.ac.uk/user/tong/statphys/sp.pdf}.

\bibitem[{Laplace(1774)}]{Laplace1774}
Laplace PS.
\newblock Memoir on the probability of causes of events.
\newblock {\em Math{\'e}matique et de Physique,Tome Sixi{\'e}me\/}  (1774)
  621--656.

\bibitem[{Laplace(1986)}]{Laplace1986}
Laplace P.
\newblock Memoir of the probability of causes of events.
\newblock {\em Statistical Science\/} {\bf 1} (1986) 365--378.
\newblock Translation to English by S. M. Stigler.

\bibitem[{Press et~al.(2007)Press, Teukolsky, Vetterling, and Flannery}]{press}
Press WH, Teukolsky SA, Vetterling WT, Flannery BP.
\newblock {\em Numerical Recipes: The Art of Scientific Computing, Third
  Edition\/} (Cambridge University Press) (2007).

\bibitem[{Johnston and Wu(1995)}]{jwu}
Johnston D, Wu SMS.
\newblock {\em Foundations of Cellular Neurophysiology\/} (Bradford Books, MIT
  Press) (1995).

\bibitem[{Sterratt et~al.(2011)Sterratt, Graham, Gillies, and
  Willshaw}]{willshaw}
Sterratt D, Graham B, Gillies A, Willshaw D.
\newblock {\em Principles of Computational Modelling in Neuroscience\/}
  (Cambridge University Press) (2011), 390 .

\bibitem[{Abarbanel et~al.(2018)Abarbanel, Rozdeba, and Shirman}]{abar18}
Abarbanel HDI, Rozdeba PJ, Shirman S.
\newblock Machine learning: Deepest learning as statistical data assimilation
  problems.
\newblock {\em Neural Computation\/} {\bf 30} (2018) 2025--2055.

\bibitem[{Fano(1961)}]{fano}
Fano RM.
\newblock {\em Transmission of Information; A Statistical Theory of
  Communication\/} (MIT Press) (1961).

\bibitem[{Kostuk et~al.(2012)Kostuk, Toth, Meliza, Margoliash, and
  Abarbanel}]{biocyb2}
Kostuk M, Toth BA, Meliza CD, Margoliash D, Abarbanel HDI.
\newblock Dynamical estimation of neuron and network properties ii: path
  integral monte carlo methods.
\newblock {\em Biological Cybernetics\/} {\bf 106} (2012) 155--167.

\bibitem[{Neal(2011)}]{neal2012}
Neal R.
\newblock Mcmc using hamiltonian dynamics.
\newblock Gelman A, Jones G, Meng XL, editors, {\em Handbook of Markov Chain
  Monte Carlo\/} (Chapman and Hall; CRC Press), chap.~5 (2011).

\bibitem[{Murty and Kabadi(1987)}]{murty87}
Murty KG, Kabadi SN.
\newblock Some np-complete problems in quadratic and nonlinear programming.
\newblock {\em Mathematical Programming\/} {\bf 39} (1987) 117--129.

\bibitem[{Ye et~al.(2015{\natexlab{a}})Ye, Kadakia, Rozdeba, Abarbanel, and
  Quinn}]{Ye-et-al}
Ye J, Kadakia N, Rozdeba PJ, Abarbanel HDI, Quinn JC.
\newblock Improved variational methods in statistical data assimilation.
\newblock {\em Nonlinear Processes in Geophysics\/} {\bf 22}
  (2015{\natexlab{a}}) 205--213.
\newblock \doi{10.5194/npg-22-205-2015}.

\bibitem[{Ye et~al.(2015{\natexlab{b}})Ye, Rey, Kadakia, Eldridge, Morone,
  Rozdeba et~al.}]{Ye2015}
Ye J, Rey D, Kadakia N, Eldridge M, Morone UI, Rozdeba P, et~al.
\newblock Systematic variational method for statistical nonlinear state and
  parameter estimation.
\newblock {\em Phys. Rev. E\/} {\bf 92} (2015{\natexlab{b}}) 052901.
\newblock \doi{10.1103/PhysevE.92.052901}.

\bibitem[{Ye(2016)}]{ye}
Ye J.
\newblock {\em Systematic Annealing Approach for Statistical Data
  Assimilation\/}.
\newblock Ph.D. thesis, University of California San Diego (2016).

\bibitem[{Quinn(2010)}]{quinn}
Quinn JC.
\newblock {\em A path integral approach to data assimilation in stochastic
  nonlinear systems\/}.
\newblock Ph.D. thesis, University of California San Diego (2010).

\bibitem[{Shirman(2018)}]{sasha}
Shirman S.
\newblock {\em Strategic Monte Carlo and Variational Methods in Statistical
  Data Assimilation for Nonlinear Dynamical Systems\/}.
\newblock Ph.D. thesis, University of California San Diego (2018).

\bibitem[{Anthes(1974)}]{anthes1974}
Anthes R.
\newblock Data assimilation and initialization of hurricane prediction models.
\newblock {\em J. Atmos. Sci.\/} {\bf 31} (1974).

\bibitem[{Gelfand and Fomin(1963)}]{gfomin}
Gelfand IM, Fomin SV.
\newblock {\em Calculus of Variations\/} (Dover Publications, Inc.) (1963).

\bibitem[{{Metropolis} et~al.(1953){Metropolis}, {Rosenbluth}, {Rosenbluth},
  {Teller}, and {Teller}}]{MC}
{Metropolis} N, {Rosenbluth} AW, {Rosenbluth} MN, {Teller} AH, {Teller} E.
\newblock {Equation of State Calculations by Fast Computing Machines}.
\newblock {\em J. Chem. Phys.\/} {\bf 21} (1953) 1087--1092.
\newblock \doi{10.1063/1.1699114}.

\bibitem[{Wong et~al.(2018)Wong, Hao, and Abarbanel}]{wong2018}
Wong AS, Hao K, Abarbanel HDI.
\newblock Precision annealing for monte carlo methods: Application to the
  shallow water equations  (2018).
\newblock In Preparation, 2018.

\bibitem[{Doupe and Kuhl(1999)}]{DoupeKuhl99}
Doupe AJ, Kuhl PK.
\newblock Birdsong and human speech: Common themes and mechanisms.
\newblock {\em Annual Review of Neuroscience\/} {\bf 22} (1999) 567--631.
\newblock \doi{10.1146/annurev.neuro.22.1.567}.
\newblock PMID: 10202549.

\bibitem[{Daou et~al.(2013)Daou, Ross, Johnson, Hyson, and Bertram}]{Daou2013}
Daou A, Ross M, Johnson F, Hyson R, Bertram R.
\newblock Electrophysiological characterization and computational models of hvc
  neurons in the zebra finch.
\newblock {\em Journal of Neurophysiology\/} {\bf 110} (2013) 1227--1245.

\bibitem[{Bolhuis et~al.(2010)Bolhuis, Okanoya, and Scarff}]{natrevneuro11}
Bolhuis JJ, Okanoya K, Scarff C.
\newblock Twitter evolution: Converging mechanisms in birdsong and human
  speech.
\newblock {\em Nature Reviews Neuroscience\/} {\bf 11} (2010) 747--759.

\bibitem[{Fee and Scharff(2010)}]{songbirdlearning}
Fee MS, Scharff C.
\newblock The songbird as a model for the generation and learning of complex
  sequential behaviors.
\newblock {\em ILAR Journal\/} {\bf 51} (2010) 362--377.
\newblock \doi{10.1093/ilar.51.4.362}.

\bibitem[{Mooney(2009)}]{Mooney2009}
Mooney R.
\newblock Neuronal mechanisms for learned birdsong.
\newblock {\em Learning and Memory\/} {\bf 16} (2009) 655--669.
\newblock \doi{10.1101/lm.1065209}.

\bibitem[{Simonyan et~al.(2012)Simonyan, Horwitz, and Jarvis}]{Simonyan2012}
Simonyan K, Horwitz B, Jarvis E.
\newblock Dopamine regulation of human speech and bird song: A critical review.
\newblock {\em Brain and Language\/} {\bf 122} (2012) 142--150.
\newblock \doi{10.1016/j.bandl.2011.12.009}.

\bibitem[{Teramitsu et~al.(2004)Teramitsu, Kudo, London, Geschwind, and
  White}]{Teramitsu2004}
Teramitsu I, Kudo LC, London SE, Geschwind DH, White SA.
\newblock Parallel foxp1 and foxp2 expression in songbird and human brain
  predicts functional interaction.
\newblock {\em Journal of Neuroscience\/} {\bf 24} (2004) 3152--3163.
\newblock \doi{10.1523/JNEUROSCI.5589-03.2004}.

\bibitem[{Jarvis et~al.(2005)Jarvis, G{\"u}nt{\"u}rk{\"u}n, Bruce, Csillag,
  Karten, Kuenzel et~al.}]{Jarvis2005}
Jarvis ED, G{\"u}nt{\"u}rk{\"u}n O, Bruce L, Csillag A, Karten H, Kuenzel W,
  et~al.
\newblock Avian brains and a new understanding of vertebrate brain evolution.
\newblock {\em Nature Review Neuroscience\/} {\bf 6} (2005) 151--159.
\newblock \doi{10.1038/nrnl1606}.

\bibitem[{Doupe et~al.(2005)Doupe, Perkel, Reiner, and Stern}]{Doupe2005}
Doupe AJ, Perkel DJ, Reiner A, Stern EA.
\newblock Birdbrains could teach basal ganglia research a new song.
\newblock {\em Trends in Neuroscience\/} {\bf 28} (2005) 353--363.
\newblock \doi{10.1016/j.tins.2005.05.005}.

\bibitem[{Nottebohm et~al.(1976)Nottebohm, Stokes, and Leonard}]{Nottebohm76}
Nottebohm F, Stokes TM, Leonard CM.
\newblock Central control of song in the canary, serinus canarius.
\newblock {\em Journal of Comparative Neurology\/} {\bf 165} (1976) 457--486.
\newblock \doi{10.1002/cne.901650405}.

\bibitem[{Fortune and Margoliash(1995)}]{Margoliash95}
Fortune ES, Margoliash D.
\newblock Parallel pathways and convergence onto hvc and adjacent neostriatum
  of adult zebra finches (taeniopygia guttata).
\newblock {\em Journal of Comparative Neurology\/} {\bf 360} (1995) 413--441.
\newblock \doi{10.1002/cne.903600305}.

\bibitem[{Brainard and Doupe(2000)}]{doupe2000}
Brainard MS, Doupe AJ.
\newblock Auditory feedback in learning and maintenance of vocal behaviour.
\newblock {\em Nature Reviews Neuroscience\/} {\bf 1} (2000).

\bibitem[{Hamill et~al.(1981)Hamill, Marty, Neher, Sakmann, and
  Sigworth}]{Hamill1981}
Hamill OP, Marty A, Neher E, Sakmann B, Sigworth FJ.
\newblock Improved patch-clamp techniques for high-resolution current recording
  from cells and cell-free membrane patches.
\newblock {\em Pfl{\"u}gers Archiv\/} {\bf 391} (1981) 85--100.
\newblock \doi{10.1007/BF00656997}.

\bibitem[{Hern{\'a}ndez-Ochoa and Schneider(2012)}]{Hernandez2012}
Hern{\'a}ndez-Ochoa EO, Schneider MF.
\newblock Voltage clamp methods for the study of membrane currents and sr
  ca$^{2+}$ release in adult skeletal muscle fibres.
\newblock {\em Progress in Biophysics and Molecular biology\/} {\bf 108} (2012)
  98--118.

\bibitem[{Bagal et~al.(2015)Bagal, Marron, Owen, Storer, and Swain}]{Bagal2015}
Bagal SK, Marron BE, Owen RM, Storer RI, Swain NA.
\newblock Voltaged gated sodium channels as drug discovery targets.
\newblock {\em Channels (Austin)\/} {\bf 9} (2015) 360--366.
\newblock \doi{10.1080/19336950.2015.1079674}.

\bibitem[{Frolov and Weckstr{\"o}m(2016)}]{frolov}
Frolov RV, Weckstr{\"o}m M.
\newblock Ion channels as therapeutic targets, part a.
\newblock {\em Advances in Protein Chemistry and Structural Biology\/} {\bf
  103} (2016) 1--386.

\bibitem[{Daou and Margoliash(2018)}]{margdaou}
Daou A, Margoliash D.
\newblock Things to say about birds; brothers and not.
\newblock {\em Waiting for Godot\/} {\bf 11901} (2018).

\bibitem[{Breen(2017)}]{breen}
Breen D.
\newblock {\em Characterizing Real World Neural Systems Using Variational
  Methods of Data Assimilation\/}.
\newblock Ph.D. thesis, University of California San Diego (2017).

\bibitem[{Wang et~al.(2017)Wang, Breen, Akinin, Broccard, Abarbanel, and
  Cauwenberghs}]{VLSI}
Wang J, Breen D, Akinin A, Broccard F, Abarbanel HDI, Cauwenberghs G.
\newblock Assimilation of biophysical neuronal dynamics in neuromorphic vlsi.
\newblock {\em Biomedical Circuits and Systems\/} {\bf 11} (2017) 1258--1270.

\bibitem[{Kadakia et~al.(2016)Kadakia, Armstrong, Breen, Daou, Margoliash, and
  Abarbanel}]{Nirag2016}
Kadakia N, Armstrong E, Breen D, Daou A, Margoliash D, Abarbanel H.
\newblock Nonlinear statistical data assimilation for hvcra neurons in the
  avian song system.
\newblock {\em Biological Cybernetics\/} {\bf 110} (2016) 417--434.

\bibitem[{Breen et~al.(2016)Breen, Shirman, Armstrong, Kadakia, and
  Abarbanel}]{Breen2016HVCI}
Breen D, Shirman S, Armstrong E, Kadakia N, Abarbanel H.
\newblock Hvci neuron properties from statistical data assimilation (2016).

\bibitem[{Nogaret et~al.(2016)Nogaret, Meliza, Margoiliash, and
  Abarbanel}]{scirept2016}
Nogaret A, Meliza CD, Margoiliash D, Abarbanel.
\newblock Automatic construction of predictive neuron models through large
  scale assimilation of electrophysiological data.
\newblock {\em Scientific Reports\/} {\bf 6} (2016).
\newblock \doi{10.1038/srep32749}.

\bibitem[{Mooney and Prather(2005)}]{Mooney2005}
Mooney R, Prather JF.
\newblock The hvc microcircuit: The synaptic basis for interactions between
  song motor and vocal plasticity pathways.
\newblock {\em Journal of Neuroscience\/} {\bf 25} (2005) 1952--1964.
\newblock \doi{10.1523/JNEUROSCI.3726-04.2005}.

\bibitem[{Kosche et~al.(2015)Kosche, Vallentin, and Long}]{Kosche15}
Kosche G, Vallentin D, Long MA.
\newblock Interplay of inhibition and excitation shapes a premotor neural
  sequence.
\newblock {\em J Neurosci.\/} {\bf 35} (2015) 1217--1227.

\bibitem[{Goldman(1943)}]{Goldman43}
Goldman DE.
\newblock Potential, impedance, and rectification in membranes.
\newblock {\em The Journal of General Physiology\/} {\bf 27} (1943) 37--60.
\newblock \doi{10.1085/jgp.27.1.37}.

\bibitem[{Dutar et~al.(1998)Dutar, Vu, and Perkel}]{Dutar98}
Dutar P, Vu HM, Perkel DJ.
\newblock Multiple cell types distinguished by physiological, pharmacological,
  and anatomic properties in nucleus hvc of the adult zebra finch.
\newblock {\em Journal of Neurophysiology\/} {\bf 80} (1998) 1828--1838.
\newblock \doi{10.1152/jn.1998.80.4.1828}.
\newblock PMID: 9772242.

\bibitem[{Mooney(2000)}]{Mooney2000}
Mooney R.
\newblock Different subthreshold mechanisms underlie song selectivity in
  identified hvc neurons of the zebra finch.
\newblock {\em Journal of Neuroscience\/} {\bf 20} (2000) 5420--5436.
\newblock \doi{10.1523/JNEUROSCI.20-14-05420.2000}.

\bibitem[{Kubota and Taniguchi(1998)}]{KubotaTaniguchi98}
Kubota M, Taniguchi I.
\newblock Electrophysiological characteristics of classes of neuron in the hvc
  of the zebra finch.
\newblock {\em Journal of Neurophysiology\/} {\bf 80} (1998) 914--923.
\newblock \doi{10.1152/jn.1998.80.2.914}.
\newblock PMID: 9705478.

\end{thebibliography}

\end{document}